\documentclass[twocolumn,prd,aps,amsmath,floatfix]{revtex4} 

\usepackage{epsfig} 
\newcommand{\bea}{\begin{eqnarray}} 
\newcommand{\eea}{\end{eqnarray}}   
\font\cmss=cmss12  
\def\1{\hbox{{1}\kern-.25em\hbox{l}}} 
\def\bfZ{\relax{\hbox{\cmss Z\kern-.4em Z}}} 
 
\begin{document} 
\def\lsim{\mathrel{\rlap{\lower4pt\hbox{\hskip1pt$\sim$}}
    \raise1pt\hbox{$<$}}}                
\def\gsim{\mathrel{\rlap{\lower4pt\hbox{\hskip1pt$\sim$}}
    \raise1pt\hbox{$>$}}}                

\title{Modeling generalized parton distributions to describe deeply virtual Compton scattering data} 
\author{A.~Freund}
\affiliation{Institut f{\"u}r Theoretische Physik, Universit{\"a}t Regensburg,  
D-93040 Regensburg, Germany} 
\author{M.~McDermott}
\affiliation{Division of Theoretical Physics, Dept. Math. Sciences, University of Liverpool, Liverpool, L69 3BX, UK}
\author{M.~Strikman}
\affiliation{The Pennsylvania State University, Department of Physics, University Park, 16802 PA, USA} 
\medskip 

\begin{abstract} 
  We present a new model for generalized parton distributions (GPDs),
  based on the aligned jet model, which successfully describes the
  deeply virtual Compton scattering (DVCS) data from H1, ZEUS, HERMES
  and CLAS. We also present an easily implementable and flexible
  algorithm for their construction.  This new model is necessary since
  the most widely used models for GPDs, which are based on factorized double
  distributions, cannot, in their current form, describe the DVCS data
  when employed in a full QCD analysis.  We demonstrate explicitly the
  reason for the shortcoming in the data description.  We also
  highlight several non-perturbative input parameters which could be
  used to tune the GPDs, and the $t$-dependence, to the DVCS data
  using a fitting procedure.
\end{abstract} 
\maketitle 

\section{Introduction} 
 
Generalized parton distributions (GPDs) have been studied extensively
in recent years
\cite{mrgdh,ji,rad1,die,mfort,vgg,jcaf,ffgs,ffs,gb1,gb2,bemuothers}. This
interest was spurred by the realization that these distributions are
not only the basic, non-perturbative ingredients in hard, exclusive
processes such as deeply virtual Compton scattering (DVCS), or
exclusive vector meson production, but that they are generalizations
of the well known parton distribution functions (PDFs) from inclusive
reactions. GPDs incorporate both a partonic and distributional
amplitude behavior and hence contain more information about the
hadronic degrees of freedom than PDFs. In fact, GPDs are true
two-parton correlation functions, allowing access to the highly
non-trivial parton correlations inside hadrons \cite{foot}.

GPDs can be broadly characterized by the following features:
\begin{itemize}
  
\item They depend on two momentum fraction variables, a partonic
  variable defined with respect either to the incoming or the average
  of the incoming and outgoing proton momentum and the {\it
    skewedness} (which is the difference between the momentum
  fractions of two adjacent partons in the parton ladder).

\item For fixed skewedness, they are continuous functions of the
  dependent variable and span two distinct regions, the DGLAP region
  and the ERBL region, in which their evolution in scale obeys
  generalized versions of the DGLAP and ERBL evolution equations,
  respectively, and in which their behavior is qualitatively
  different.
  
\item They are even functions of the skewedness variable and the
  singlet, non-singlet and gluon distributions are either symmetric or
  anti-symmetric about the center point of the ERBL region (the
  symmetry obeyed depends on the precise definitions used).

\item The Lorentz structure of their definitions implies a polynomiality 
condition \cite{ji,rad1,gb2,poly}: their ($N-1$)-th moments are polynomials in square 
of the skewedness of degree no greater than $N/2$.  

\item  They reduce to the ordinary PDFs in the limit of zero skewedness 
(the `forward limit').

\end{itemize}
\noindent All of the above features have to be preserved under evolution in scale.

Any suggested model of GPDs should adhere to these mathematical
features.  In \cite{rad2} such a model, based on double distributions
(DDs), was suggested for the GPD input distributions (see also
\cite{rad3}).  In \cite{poly} it was realized that an additional term,
the so-called D-term, was required in the ERBL region for the
unpolarized quark singlet and gluon distributions in order to satisfy
polynomiality for even $N$.  The use of factorized \cite{factor} DDs
augmented with a D-term has become a popular phenomenological model.
Unfortunately, when this type of model for input GPDs was used in its
current form to calculate deeply virtual Compton scattering at both
leading (LO) and next-to-leading order (NLO), the results were not in
agreement with the H1 data \cite{h1} on the DVCS photon level cross
section, $\sigma(\gamma^* p \to \gamma p)$, and the HERMES and CLAS data
\cite{hermclas} on the DVCS single spin asymmetry or charge asymmetry
\cite{afmmshort,afmmlong,bemuk1}.

Another popular model for input GPDs, inspired by the aligned jet
model (AJM) \cite{ajm} and its QCD extension \cite{fs88} is based on
the observation that at a scale $Q^2\sim 1-2~\mbox{GeV}^2$ and a wide
range of $x_{bj}$, soft physics gives the dominant contribution to the
parton densities. As a result the effect of skewedness at small
$x_{bj}$ should be rather small and hence at the input scale, it is a
good approximation to set the GPDs equal to the forward PDFs at the
same parton fraction, $X$, defined with respect to the {\it incoming}
proton \cite{ffgs} (for any skewedness).  This has the advantage that
it automatically satisfies the requirements of polynomiality for the
first two moments, however one encounters infinities in the quark
singlet GPD in the middle of the ERBL region.

Another `forward model', which may be considered to be an extreme case
of a DD model, was adopted in \cite{gb2} where one assumes that the
GPD is equal to the forward PDF at the same parton momentum fraction,
$v$, with respect to the average of the {\it incoming and outgoing}
proton momentum ${\bar P} = (p + p')/2$ which implicitly contains the
skewedness.  This translates to an $X$ which is shifted to lower
values by an amount controlled by the skewedness.  This ansatz works
fine for the DGLAP region. Unfortunately in the ERBL region it also
involves sampling the forward PDFs right the way down to zero in
momentum fraction where they have not yet been measured (this is
especially problematic for singular quark distributions). In this
paper we construct an alternative, finite `forward model' for the
input GPDs, using the forward input PDFs in the DGLAP region and
imposing a simple form in the ERBL region that has the correct
symmetries and ensures polynomiality is respected in the first two
moments (see \cite{gb2} for alternative ways of dealing with this
problem). As we will demonstrate this forward model reproduces the
available data on DVCS reasonably well.

\begin{figure} 
\centering 
\mbox{\epsfig{file=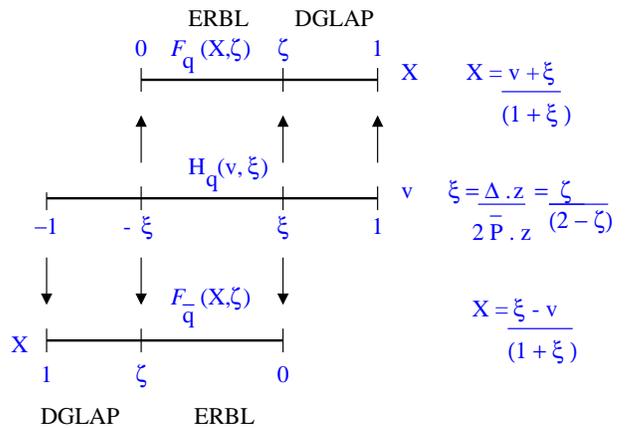,width=8.0cm,height=5.5cm}} 
\caption{The relationship between ${\cal F}^q (X,\zeta)$, ${\cal F}^{\bar q} (X,\zeta) $ and Ji's function $H^q (v,\xi)$ with $v \in [-1,1]$ and $X \in [0,1]$.} 
\label{figfdd} 
\end{figure}

Throughout this paper we will use the off-diagonal representation of
GPDs, ${\cal F}^i (X,\zeta)$, defined by Golec-Biernat and Martin
\cite{gb1} and used, for example, in the numerical solution of the GPD
evolution equations in \cite{afmmgpd} (see \cite{gb1,bemuevolve,radmus}
for other approaches to numerical evolution). They depend on the
momentum fraction $X \in [0,1]$ of the incoming proton's momentum, $p$,
and the skewedness variable $\zeta = \Delta^+/p^+$ (so that $\zeta
= x_{bj}$ for DVCS). For the quark case, the relationship of the quark
and anti-quark distributions, ${\cal F}^q (X,\zeta), {\cal F}^{\bar q}
(X,\zeta) $, to Ji's GPD $H^q (v,\xi)$ is shown in Fig.~\ref{figfdd} with $\xi=\Delta^+/2{\bar P}^+$.  More
explicitly, for $v \in [-\xi,1]$:
\begin{equation}
{\cal F}^{q,a} \left(X = \frac{v+\xi}{1 + \xi},\zeta\right) = \frac{H^{q,a} (v,\xi)}{1-\zeta/2} \, ,
\label{curlyq}
\end{equation}
\noindent and for $v \in [-1,\xi]$
\begin{equation}
{\cal F}^{{\bar q},a} \left(X = \frac{\xi -v}{1 + \xi},\zeta\right) = -\frac{H^{q,a} (v,\xi)}{1-\zeta/2} \, .
\label{curlyqbar}
\end{equation}
The two distinct transformations between $v$ and $X$ for the quark and
anti-quark cases are shown explicitly on the left hand side of
Eqs.(\ref{curlyq}, \ref{curlyqbar}) .  There are two distinct regions:
the DGLAP region, $X > \zeta$ ($|v| > \xi$), and the ERBL region, $X<\zeta$
($|v| < \xi$).  In the ERBL region, due to the fermion symmetry, ${\cal
  F}^q$ and ${\cal F}^{\bar q}$ are not independent. In fact ${\cal
  F}^{q} (X,\zeta) = -{\cal F}^{\bar q} (\zeta-X,\zeta)$, which leads to an
anti-symmetry of the unpolarized quark singlet distributions (summed
over flavor $a$), ${\cal F}^S = \Sigma_a {\cal F}^{q,a} + {\cal F}^{{\bar
    q},a}$, about the point $\zeta/2$ (the non-singlet and the gluon,
${\cal F}^g$, which is built from $v H^g_{Ji} (v,\xi)$, are symmetric
about this point). ${\cal F}^{NS,g}$ are constructed analogously to
${\cal F}^S$ from $H^{NS,g}$.

The rest of the paper is organized as follows.
Section~\ref{ddfailure} contains a detailed explanation of why
DD-based models in their current form cannot describe the data. In
Section~\ref{forward} we construct our alternative forward model for
input GPDs, which is motivated by the AJM \cite{ajm,fs88} and
describes the data well. In Section~\ref{slope} we propose a
phenomenological model for the slope of the $t$-dependence in which
the slope parameter is allowed to change with photon virtuality, $q^2
= -Q^2$. The model improves the theoretical description of the
$Q^2$-dependence of the HERA data, relative to using a constant slope.
Finally we summarize our findings in Section~\ref{con}.

\section{The problem with double distribution models}
\label{ddfailure}

In this section we discuss factorized double distribution (DD) based
models for GPDs and explain in detail why the sampling of the forward
PDF at extremely small $x$ in constructing the GPD leads to a problem
in the quark singlet GPD.

\begin{figure} 
\centering 
\mbox{\epsfig{file=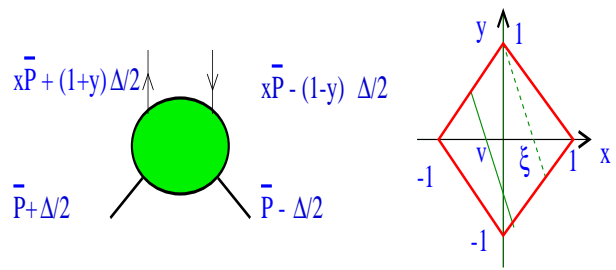,width=8cm,height=3.5cm}} 
\caption{Symmetric double distributions (left), indicating momentum fractions of the outgoing and returning partons, and (right) their physical domain.} 
\label{figdds} 
\end{figure} 

Symmetric DDs, $F_{DD}(x,y,t,Q^2)$, were introduced in \cite{rad1}
with plus momentum fractions, $x,y$, of the outgoing and returning
partons defined as shown in the left hand plot of Fig.~\ref{figdds}.
They exist on the diamond-shaped domain shown to the right of
Fig.~\ref{figdds}.  The outgoing parton lines of course only have a
single plus momentum relative to any particular external momenta, so
the GPDs are related to these DDs via a reduction integral,
involving $\delta(v - x - \xi y)$, along the off-vertical lines in the
diamond (the dotted line corresponds to $v=\xi$):
\begin{align}
&H(v,\xi) = \int^1_{-1} dx' \int^{1-|x'|}_{-1+|x'|} dy' \delta (x' + \xi y' - v) F_{DD} (x',y').
\label{reduction} 
\end{align}  

In \cite{rad2,rad3} a model for $F_{DD} (x,y,t,\mu^2)$, at the input
scale $\mu^2 = Q_0^2$, was introduced in which the functional form is
factorized in $x$ and $y$ and in the $t$-dependence, an assumption 
borne out of convenience rather than physical necessity:
\begin{equation} 
F_{DD}^{i} (x,y,\mu^2,t) = \pi^{i} (x,y) \, f^{i} (x,\mu^2) \, r^{i}(t). 
\label{DDs} 
\end{equation} 
Here $f^i$ and $r^{i}$ are the standard PDF and form factor for the
parton distribution of general type $i$. Since $t$-dependence is
assumed to factorize and thus of no importance to the following, we
will suppress it from now on.  The profile functions, $\pi^{i} (x,y)$,
are asymptotic shape functions \cite{rad1} for quarks and gluons of
the general form
\begin{equation}
\pi(x,y) = \frac{\Gamma(2b + 2)}{2^{2b+1} \Gamma^2 (b+1)} \frac{[(1 -|x|)^2 - y^2]^b}{(1 -|x|)^{2b+1}} \, , 
\label{profile}
\end{equation}
and normalized such that 
\begin{align}
\int^{1-|x|}_{-1+|x|} dy ~ \pi(x,y)= 1 \, .
\label{profile2}
\end{align}
Note that $\pi$ is an even function of both of its arguments
\cite{footpi}.  The power $b$ controls the size of the skewing effects
in the input GPD.  Usually $b=1$ is chosen for the quarks,
corresponding to maximum skewedness, whereas $b = 2$ is chosen for the
gluons. In the limit $b \to \infty$ there is no external skewedness effect,
however, since in this limit, $H(v,\xi) = q(v)$, this translates into
internal dependence on the skewedness ${\cal F}(X,\zeta) =
q\left(\frac{X-\zeta/2}{1-\zeta/2}\right)/(1-\zeta/2)$. Note that a
consequence of the above model is a ratio of GPD/PDF for the quark
singlet in the DGLAP region at $X=\zeta=x_{bj}$, which is substantially
larger than $1$ \cite{afmmgpd,radmus} for all experimentally relevant
values of $x_{bj}$.

Concerning the description of the data using the above model, it was
shown in \cite{afmmlong} that maximal skewing ($b=1$) at
`conventional' input scales ($Q_0 = 1, 2~\mbox{GeV}$) overshoots the
H1 data by a factor of $6-10$.  It was also demonstrated that one can
describe, in LO only, the H1 data without including skewedness effects
at the input scale, $Q_0 = 2~\mbox{GeV}$, if one neglects evolution
\cite{bemuk1}.  This simplification, however, is not warranted since
we know that the effects of skewed evolution are much stronger in the
region of $X \sim \zeta$ compared to forward evolution \cite{ffgs,afmmgpd}.
Note that this region strongly influences the cross section at small
$x_{bj}$ and some asymmetries both at small and large $x_{bj}$
\cite{foot3}, if they are dominated by the imaginary part of the
amplitude which is proportional to ${\cal F}^S(\zeta,\zeta)$ at LO (see for
example \cite{afmmlong} as well as \cite{bemu2,afmmamp,bemus}).  

One may wonder whether one can come closer to the data by choosing a
very low input scale and valence-like partons as in the GRV scenario
\cite{grv}, generating the rise of the parton distributions entirely
through evolution.  It turns out that choosing the canonical value of
$b_q = 1$ and GRV98 input distributions ($Q_0 = 0.51, 0.63$~GeV in LO
and NLO, respectively) the curves still overshoot the data
considerably.  Indeed, even if one tries to minimize the effect of the
enhancement due to skewedness at the input scale, by choosing a large
value of $b_q =100$, evolution still drives the prediction above the
data by at least a factor of $\sim4$.  Since, for example, $d\sigma_{DVCS}\simeq
|{\cal F}^S(\zeta,\zeta)|^2$, the large enhancement of the quark singlet GPD
at $X=\zeta$ is the root of the problem in this model.  As we will explain
below the origin of the enhancement stems from sampling singular
forward sea distributions at extremely small $x$ in the DD-based
model. To understand the last statement, one has to first establish
the regions in $x$ in which the PDFs are sampled in the double
distribution model, particularly at small $x$.

First, having defined the model for the factorized double distribution
in Eq.~(\ref{DDs}) one may then perform the $y'$-integration in
Eq.~(\ref{reduction}) using the delta function.  This then modifies
the limits on the $x'$-integration according to the region concerned:
for the DGLAP region $X > \zeta ~(v > \xi)$ one has for the quark GPD

\begin{align}  
&{\cal F}^{q,a} (X,\zeta) = \nonumber\\ 
&\frac{2}{\zeta} \int^{X}_{\frac{X-\zeta}{1-\zeta}} dx'  
\pi^q \left (x', \frac{2}{\zeta} (X - x') + x' -1 \right) q^a (x') \, . 
\end{align} 
For the anti-quark GPD in the DGLAP region $X > \zeta ~(v < - \xi)$ one has 
\begin{align}  
&{\cal F}^{\bar q,a} (X,\zeta) = \nonumber\\ 
&\frac{2}{\zeta} \int^{\frac{-X+\zeta}{1-\zeta}}_{-X} dx' \pi^q \left  
(x', -\frac{2}{\zeta} (X + x') + x' +1 \right) {\bar q}^a(|x'|) \, ,
\label{DGLAPqbar}  
\end{align} 
\noindent changing variables from $x \to - x$ and exploiting the fact that the profile functions are even in both arguments one arrives at
\begin{align}  
  &{\cal F}^{\bar q,a} (X,\zeta) = \nonumber\\
  &\frac{2}{\zeta} \int_{\frac{X-\zeta}{1-\zeta}}^{X} dx' \pi^q \left( x', \frac{2}{\zeta}
    (X - x') + x' - 1 \right ){\bar q}^a(|x'|) \, ,
\end{align} 
\noindent so that the singlet and non-singlet quark distributions are given by
\begin{align}  
&{\cal F}^{S} (X,\zeta) = \sum_a {\cal F}^{q,a} + {\cal F}^{{\bar q},a} \nonumber\\ 
& \sum_a \frac{2}{\zeta} \int_{\frac{X-\zeta}{1-\zeta}}^{X} dx' \pi^q \left(x', {\tilde y}(x') \right ) ~[q^a (x') + {\bar q}^a (x')] \, , \nonumber\\
&{\cal F}^{NS,a } (X,\zeta) = {\cal F}^{q,a} - {\cal F}^{{\bar q},a} \nonumber\\ 
& \frac{2}{\zeta} \int_{\frac{X-\zeta}{1-\zeta}}^{X} dx' \pi^q \left(x', {\tilde y}(x') \right ) ~[q^a (x') - {\bar q}^a (x')] \, , 
\label{dglapsns}
\end{align} 
\noindent where ${\tilde y}(x') =  2 (X - x')/\zeta  + x' -1 $. 

In the ERBL region, $ X < \zeta ~(|v| < \xi$) integration over $y$ leads to:  
\begin{align} 
&{\cal F}^{q,a} (X,\zeta) = \nonumber \\  
&\frac{2}{\zeta} \left[\int^{X}_{0} dx' \pi^q \left(x', \frac{2}{\zeta} (X - x') + x' -1 \right) q^a (x') \, - \right. \nonumber \\  
&\left. \int^{0}_{X-\zeta} dx' \pi^q \left(x', \frac{2}{\zeta} (X - x') + x' -1 \right) {\bar q}^a (|x'|) \right] \, ,  \nonumber \\  
&{\cal F}^{{\bar q},a} (X,\zeta) = \nonumber \\  
& - \frac{2}{\zeta} \left[ \int^{\zeta - X}_{0} dx' \pi^q \left(x', -\frac{2}{\zeta} (X + x') + x' + 1 \right) q^a (x') \, - \right. \nonumber \\  
&\left. \int^{0}_{-X} dx' \pi^q \left(x', -\frac{2}{\zeta} (X + x') + x' + 1  \right) {\bar q}^a (|x'|) \right] \, .  
\label{erblqqbar1}
\end{align}
\noindent Again, using $x \to - x$ and $\pi(|x'|,|y'|)$, one gets
\begin{align} 
{\cal F}^{q,a} (X,\zeta) & = 
\frac{2}{\zeta} \Big[\int^{X}_{0} dx' \pi^q \left(x', {\tilde y}(x') \right) q^a (x')\nonumber\\
&- \int_{0}^{\zeta -X} dx' \pi^q \left(x', {\tilde y}(-x') \right) {\bar q}^a (x') \Big] \, ,  \nonumber \\  
{\cal F}^{{\bar q},a} (X,\zeta) &= -\frac{2}{\zeta} \Big[ \int^{\zeta - X}_{0} dx' \pi^q \left(x', {\tilde y}(-x') \right) q^a (x')\nonumber\\
&- \int_{0}^{X} dx' \pi^q \left(x', {\tilde y}(x') \right) {\bar q}^a (x') \Big] \, .  
\label{erblqqbar2}
\end{align}
Hence, for the singlet and non-singlet combinations one has
\begin{align}  
&{\cal F}^{S} (X,\zeta) = \sum_a {\cal F}^{q,a} + {\cal F}^{{\bar q},a} \nonumber\\ 
& \sum_a \frac{2}{\zeta}\Big[\int_{0}^{X} dx' \pi^q \left(x', {\tilde y}(x') \right ) ~[q^a (x') + {\bar q}^a (x')]\Big]\nonumber\\
&- \int_{0}^{\zeta-X} dx' \pi^q \left(x', {\tilde y}(-x') \right ) ~[q^a (x') + {\bar q}^a (x')]\Big]\, , \nonumber\\
&{\cal F}^{NS,a } (X,\zeta) = {\cal F}^{q,a} - {\cal F}^{{\bar q},a} \nonumber\\ 
& \frac{2}{\zeta} \left[ \int_{0}^{X} dx' \pi^q \left(x', {\tilde y}(x') \right ) ~[q^a (x') - {\bar q}^a (x')] \right. \nonumber\\ 
&+ \left.\int^{\zeta-X}_{0} dx' \pi^q \left(x', {\tilde y}(-x') \right ) ~[q^a (x') - {\bar q}^a (x')] \right] \, .
\label{erblsns}
\end{align} 
These expressions clearly satisfy the correct symmetries properties, i.e., $F^{S}(\zeta-X,\zeta) = - F^{S}(X,\zeta) $, $F^{NS,a}(\zeta-X,\zeta) =  F^{NS,a}(X,\zeta) $ (n.b. ${\tilde y} (x') \to {\tilde y} (-x')$ when $X \to \zeta - X$ ).
Analogously for the gluon one obtains 
\begin{align}
&{\cal F}^{g} (X,\zeta) =
\frac{2}{\zeta} \int_{\frac{X-\zeta}{1-\zeta}}^{X} dx' \pi^g \left(x', {\tilde y}(x') \right ) ~x'g(x') \, , 
\label{dglapg}
\end{align}
for the DGLAP region, and
\begin{align}
{\cal F}^{g} (X,\zeta) &=
\frac{2}{\zeta}\Big[ \int_{0}^{X} dx' \pi^g \left(x', {\tilde y}(x') \right ) ~x'g(x')\nonumber\\
&+ \int_{0}^{\zeta-X} dx' \pi^g \left(x', {\tilde y}(-x') \right ) ~x'g(x')\Big] \, ,  
\label{erblg}
\end{align}
for the ERBL region (which is symmetric under $X \to \zeta - X$).

Inspection of the integration limits in Eqs.~(\ref{dglapsns},
\ref{erblsns}) highlights the main problem. In the limit $ X \to \zeta$, as
a result of the lower limits of the integrals the forward PDF is
sampled closer and closer to $x' = 0$, where it has not yet been
measured. This will be irrelevant providing the integrand is
sufficiently non-singular in $x'$ in this region i.e. it can happen if
the profile functions, $\pi^i$, provide a strong suppression of this
region, or if the PDFs themselves are sufficiently non-singular.
However, we know that phenomenological quark and sometimes even gluon
input distributions are singular in the small $x$ region. In the quark
case this problem is made worse by the fact that we sample the number
distribution $q(x')$ rather than the momentum distribution $x' q(x')$
(so that a non-singular momentum distribution $xq(x) \propto x^{a}$ for $a \in
[0,1]$ will give a singular number distribution $q(x) \propto x^{a-1}$).

It turns out that for realistic quark distributions the region close
is $x' = 0$ is very significantly sampled for small $\zeta=x_{bj}$. This
leads to two serious problems.  Firstly, the forward distributions are
unknown here so one must extrapolate the `known' analytic forms
downwards in $x'$. Secondly, and much more importantly, it leads to a
very significant enhancement of the quark singlet GPDs relative to the
PDFs for $X \approx \zeta$, i.e., the region most relevant for DVCS. Though of
paramount importance for DVCS, this region is but a small region of
phase space where the current factorized DD models fail. 

\begin{figure}  
\centering
\mbox{\epsfig{file=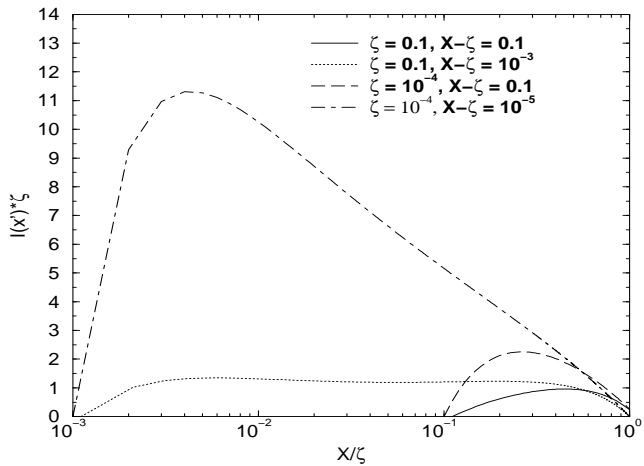,width=8.5cm,height=6.25cm}} 
\caption{The integrand of Eq.(\ref{dglapsns}), illustrating how the up singlet PDF is sampled in the DGLAP region close to the boundary of the ERBL region, to produce the up singlet GPD.}
\label{intx}
\end{figure}

\begin{figure}  
\centering
\mbox{\epsfig{file=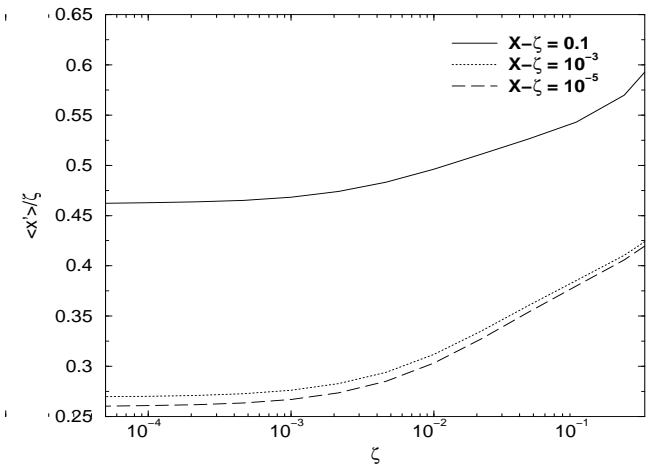,width=8.5cm,height=6.25cm}} 
\caption{The average value of $x'$ sampled in the DGLAP region in the double distribution model, for the up singlet GPD, close to the boundary with the ERBL region as a function of the skewedness. Several values of $X-\zeta$ are shown.}
\label{intx2}
\end{figure}

\begin{figure}  
\centering
\mbox{\epsfig{file=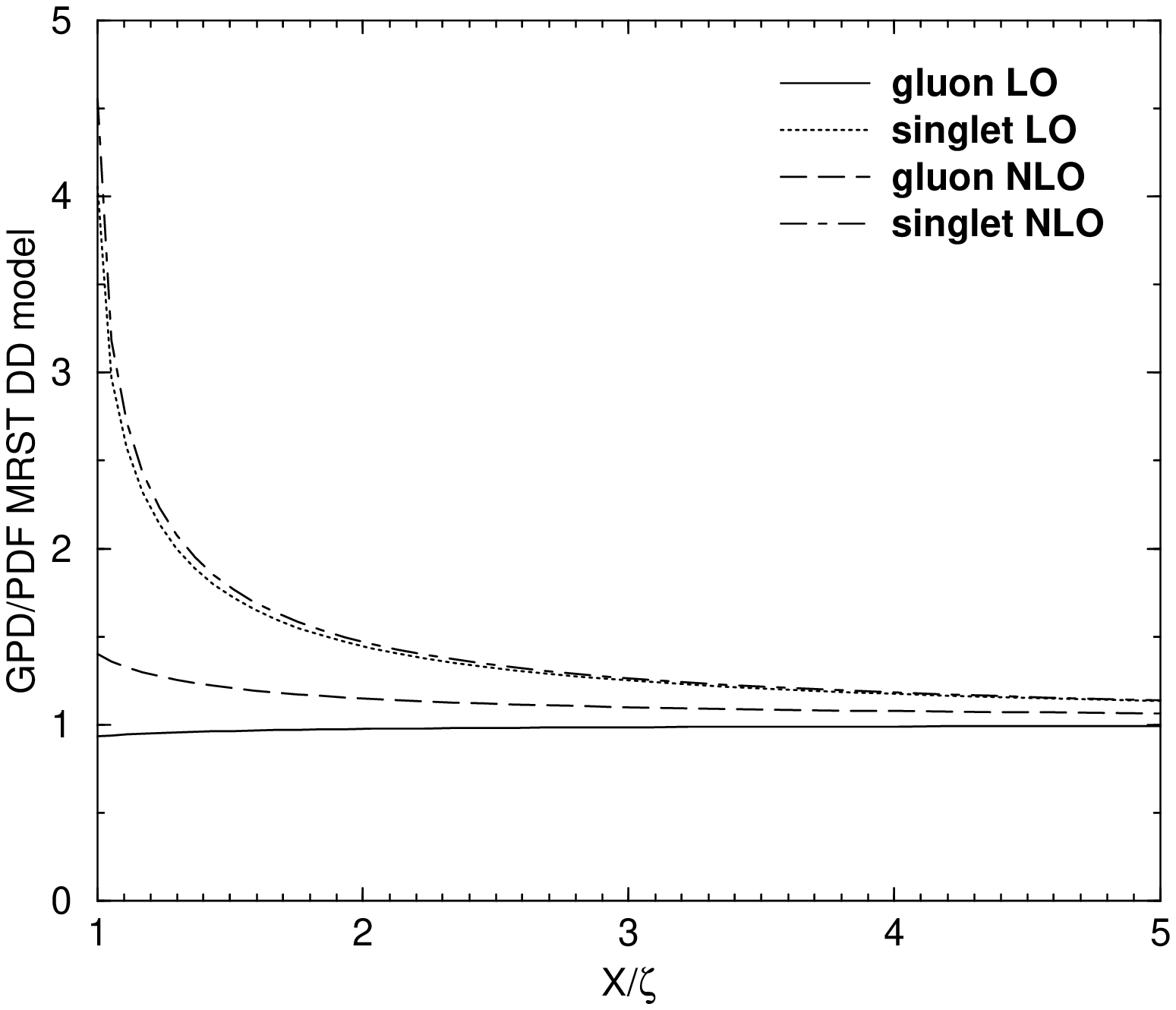,width=8.5cm,height=6.25cm}} 
\mbox{\epsfig{file=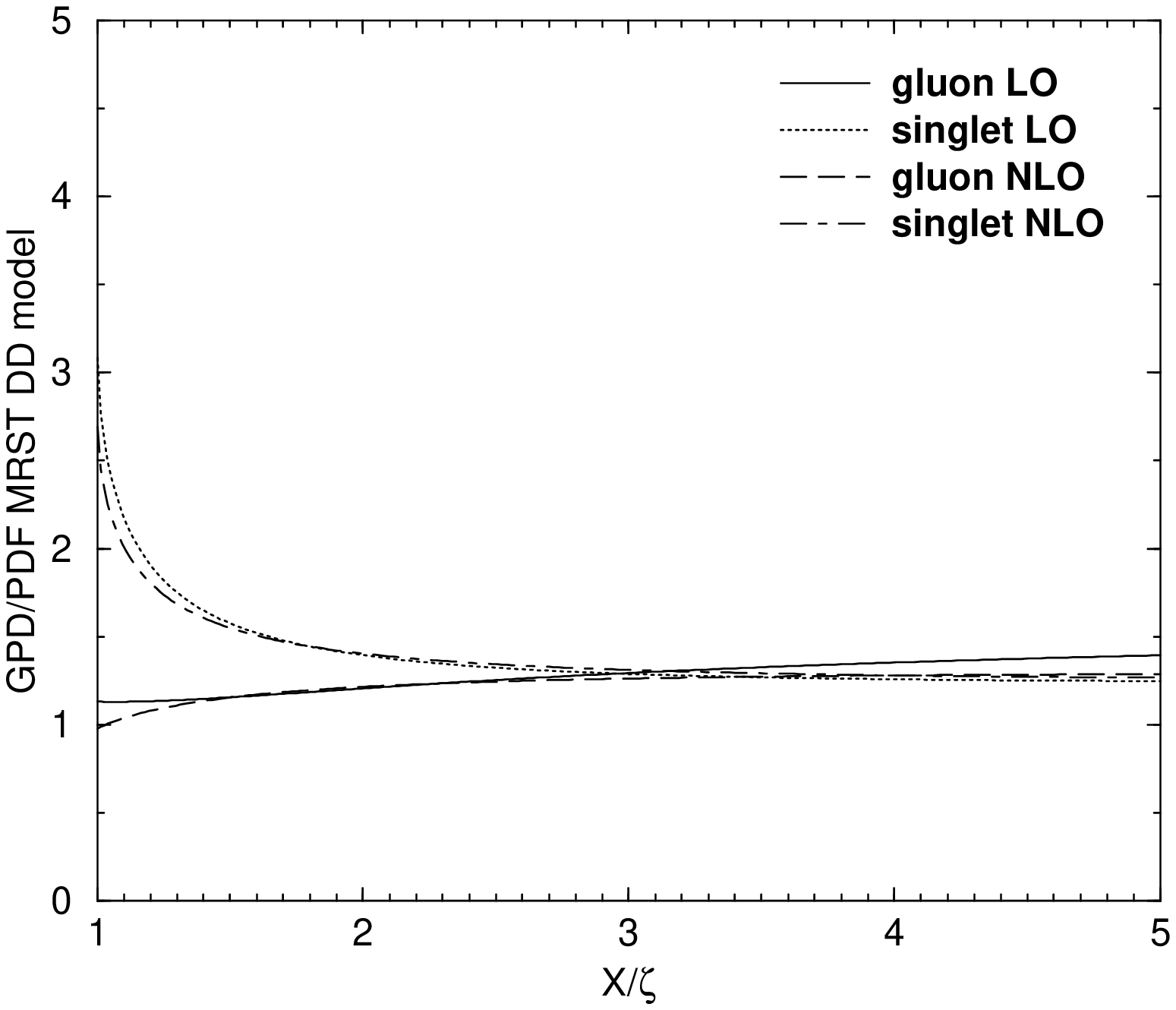,width=8.5cm,height=6.25cm}} 
\caption{The ratio GPD to PDF at $\zeta=0.0001$ (upper plot) and $\zeta=0.1$ (lower plot) for the quark singlet and gluon in the double distribution model, using MRST01 distributions in LO and NLO, at the input scale $Q_0 = 1$~GeV. Note the large enhancement of the quark singlet close to $X=\zeta$.}
\label{ratiodd}
\end{figure}

We illustrate this using a series of three figures relating to the
formation of the quark singlet GPD in the DGLAP region close to $X=\zeta$.
Fig.~\ref{intx} shows the integrand, $I(x') = \pi^u (x',{\tilde y})
~u^{S} (x')$, of Eq.(\ref{dglapsns}) for the up quark singlet
(multiplied by $\zeta$) as a function of $x'/\zeta$ for two values of $\zeta =
0.1, 0.0001$ and two values of $X-\zeta = 0.1, 0.001$. Clearly as $X$
approaches $\zeta$ the PDF is sampled at progressively smaller values of
$x' \ll \zeta$, where for small $\zeta$ it is unknown.  Fig.~\ref{intx2} shows
the average value of $x'$ sampled in this integral (divided by $\zeta$) as
a function of $\zeta$ for several values of $X-\zeta$. For very small values
of $X-\zeta$ the average value of $x'$ settles down to about $\zeta/4$, for
small $\zeta$.  Finally in Fig.~\ref{ratiodd} we show, for MRST input
PDFs, the ratio of the quark singlet GPD to PDF at $\zeta = 0.0001,0.1$,
for the canonical value of the power, $b_q=1$, in Eq.~(\ref{profile}).
Note the large enhancement of the GPD at $X \approx \zeta$, particularly for
small $\zeta$ in the upper plot. We emphasize that this enhancement, which
leads to an overshoot of the DVCS data, is built in right at the start
in the modeling of the quark singlet GPD at the input scale.  One
also sees that for the gluon, which uses $x'g(x')$, and $b_g = 2$, the
ratio remains close to unity.

The most important enhancement effect in the valence region, $\zeta \gsim
0.1$, originates from the relative shift of the parton momentum
fraction $X$ to smaller values close to $X=\zeta$ (although the
enhancement from small $\langle x^{'}\rangle$ is still significant).  As we will
show in the next section, the assumption that $H(v,\xi)=(1-\zeta/2){\cal
  F}(X,\zeta) = q(v)$ with $v=(X-\zeta/2)(1-\zeta/2)$ gives a good description of
the data at both small and large $x_{bj}$. As stated before, this
corresponds to a factorized DD-model with $b=\infty$, i.e., with no external
skewedness. However, in terms of a comparison of GPD to forward PDF,
there is a residual effect of skewedness since one now has to compare
$q(v)$ with $q(X)$.  Since we are comparing number distributions which
are more singular than momentum distributions, any shift in the
momentum fraction to smaller values will lead to a quite a large
enhancement of $q(v)$ relative to $q(X)$. For CTEQ6M for example the
enhancement at $X=0.1$ and $\zeta=0.1$ is about $1.7$ for the quark
singlet, which increases further if more skewedness is added by
decreasing $b$.

However, as we will demonstrate in Section~\ref{forward}, the
available data allows little room for further enhancement due to
skewedness at the input scale since the LO result, at least, is
already close to the upper bound of the experimental errors.
Therefore, only the extremal ``$b=\infty$'' version of the current
factorized DD-model can be used to describe the data.  An obvious
solution to this is to modify the quark singlet profile functions in
Eq.(\ref{DDs}) in such a way as to suppress the region of very small
$x'$. However, one must find a new functional form which achieves this
without spoiling the known mathematical features of GPDs discussed
above. Alternatively, one abandons the attempt to model a DD using a
factorized form $\pi(x,y)f(x)$, though appealing due to its simplicity
but possibly too simplistic in its form. This remains an open problem
and has to be addressed by those who wish to use the double distribution
framework, factorized or not, to model GPDs.

\section{The forward input model and the aligned jet model}
\label{forward}

In this section we revisit the logic of setting the GPDs equal to the
forward PDFs by proposing an alternative forward model to that
suggested in \cite{gb2}, with suitably-symmetrized input GPDs in the
ERBL region constructed so as to satisfy the requirements of polynomiality
for the first two moments.
      
In \cite{ffs} DVCS was predicted to be measurable at DESY-HERA and,
allowing for the freedom associated with choosing the slope parameter,
$B$, the predictions successfully describe both the H1 data \cite{h1}
and the recent ZEUS result \cite{zeus,zeus2} on the photon-level DVCS
cross section. This was achieved by modeling the imaginary part of
the DVCS amplitude at the input scale using the aligned jet model
(AJM) \cite{fs88}.  This was then compared to the imaginary part
of the DIS amplitude, calculated within the same framework, which was
found to be smaller by a factor of about two.  The comparison enabled
the normalization of the DVCS amplitude at the input scale to be set
using $F_2$ structure function data.  The DVCS amplitude was then
evolved to higher scales using LO skewed evolution in perturbative
QCD.

\begin{figure}
\centering
\mbox{\epsfig{file=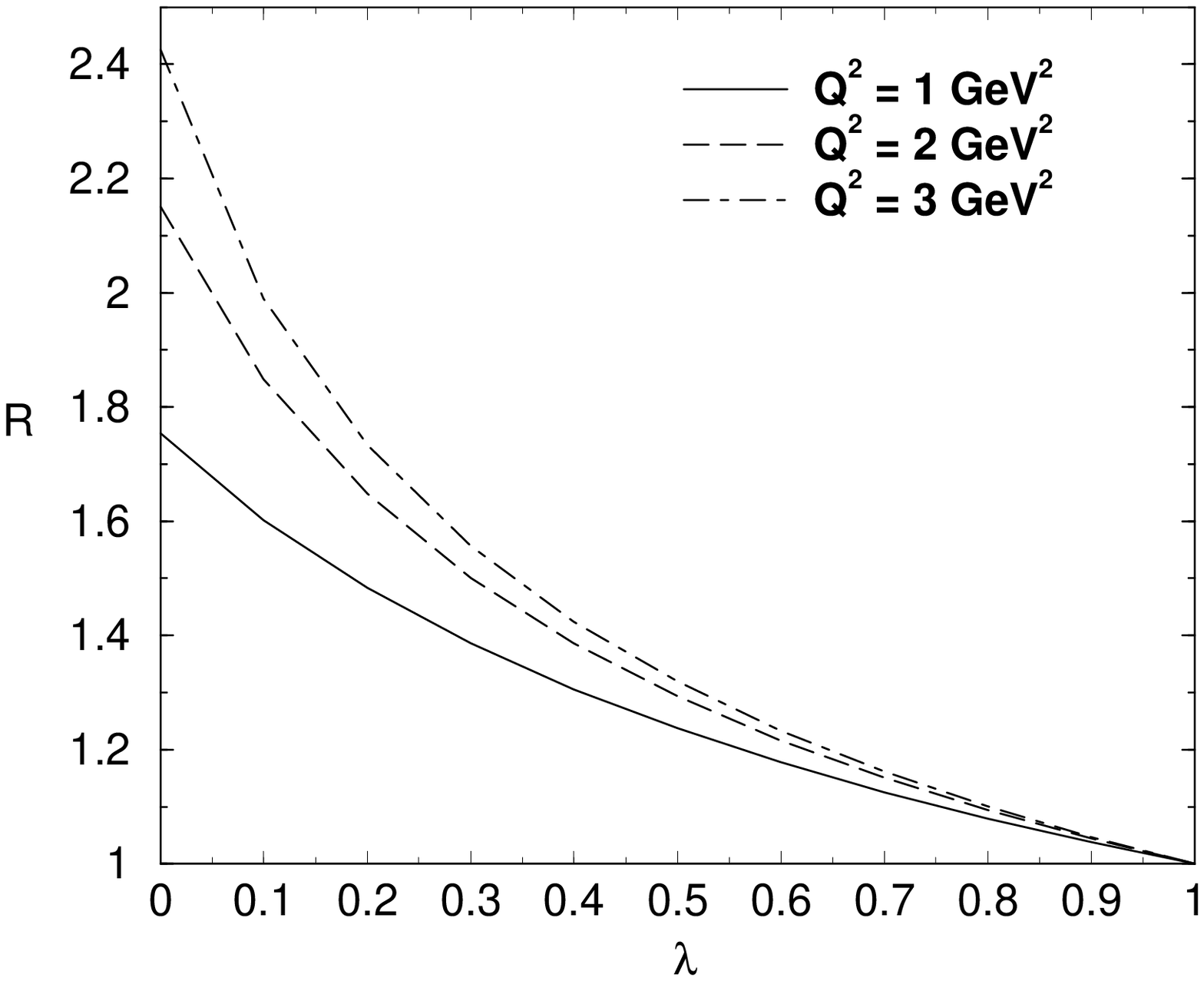,width=8.5cm,height=6.25cm}}
\mbox{\epsfig{file=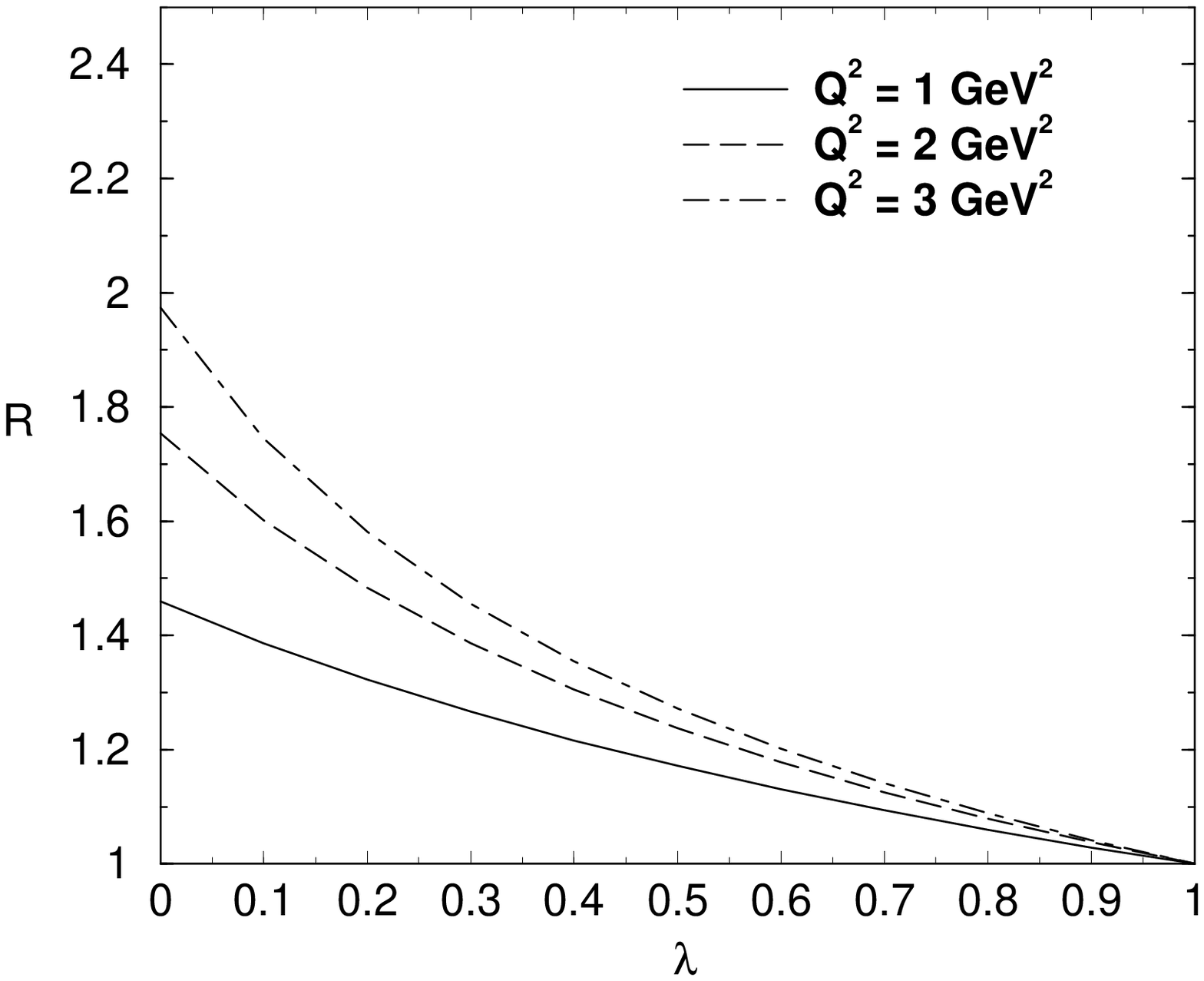,width=8.5cm,height=6.25cm}}
\caption{The ratio $R$ as a function of $\lambda = Q^{'2}/Q^2$ for several value of $Q^2$ and two values of $M_0^2 = 0.4$~GeV$^2$ (upper plot) and $0.8$~GeV$^2$ (lower plot).}
\label{ajmratio}
\end{figure}

The basic relation between the DVCS and DIS amplitudes, using the AJM,
is given by \cite{ffs}
\begin{align}
R = \frac{\mbox{Im} {\cal T}_{{\small{\mbox{DVCS}}}}}{\mbox{Im} {\cal T}_{{{\mbox{DIS}}}}} = \ln \left(1+\frac{Q^2}{M^2_0}\right)\left(1+\frac{M^2_0}{Q^2}\right) \simeq 1.5 - 2.5 \, ,
\label{ajmeq1}
\end{align}
where $Q^2$ for the AJM is typically $1 - 3$~GeV$^2$ and $M_0$ is a
hadronic scale which roughly corresponds to the lowest allowed,
excited intermediate state in the $s$-channel.  Therefore, $M^2_0\sim
0.4-0.6~\mbox{GeV}^2$, or about $m^2_{\rho}$.  The AJM neglects the
contribution of quarks with large transverse momenta in the quark loop
attached to the photons in the handbag diagram.  Since the
contribution of small transverse momenta is more symmetric than the
one at large transverse momenta, the AJM may somewhat overestimate the
effect of skewedness at the input scale.  Eq.~(\ref{ajmeq1}) can be
generalized to demonstrate how the forward limit $\mbox{Im} {\cal
  T}_{{\small{\mbox{DVCS}}}}= \mbox{Im} {\cal T}_{{{\mbox{DIS}}}}$ is
achieved, i.e., how the skewedness effect is reduced by giving the
outgoing photon a space-like virtuality, $q^{'2} = - Q^{'2}$:
\begin{align}
R = \frac{\mbox{Im} {\cal T}_{{\small{\mbox{DVCS}}}} }{\mbox{Im} {\cal T}_{{{\mbox{DIS}}}}} = \ln\left(\frac{1+\frac{Q^2}{M^2_0}}{1+\frac{Q^{'2}}{M^2_0}}\right)\frac{1+\frac{M^2_0}{Q^2}}{1-\frac{Q^{'2}}{Q^2}} \, .
\label{ajmeq2}
\end{align}
This procedure allows one to derive a very important relation between
the relative momentum fractions of the outgoing and returning partons,
$X$ and $-(X - \zeta)$ of the quark singlet GPD, and the virtualities of
the incoming and outgoing photons
\begin{equation}
\lambda = \frac{X-\zeta}{X} = \frac{Q^{'2}}{Q^2} \, .
\end{equation} 
We illustrate this point in Fig.~\ref{ajmratio} by plotting $R$ as a
function of $\lambda$ for several values of $Q^2$ and two values of $M_0^2$
to demonstrate the relative insensitivity of $R$ (within $20 \sim 30\%$)
to $M_0$.

The plot shows that as $X$ increases relative to $\zeta$ in the DGLAP
region the ratio drops rapidly to its forward limit. For example, at
$\lambda = 1/2$, i.e., $X = 2 \zeta$, the curves are very flat and there is only
a modest enhancement of $20 \sim 40 \%$.  One also encounters this
behavior in the DD model if one investigates the ratio of the GPD to
the PDF in the DGLAP region (see for example \cite{afmmgpd,radmus}).
It would therefore be advantageous to be able to directly relate $R$
to a ratio of GPD to PDF.  Trusting that perturbative QCD is
applicable at the AJM input scale one can, in LO at least where the
coefficient function is trivial, directly translate the ratio in
amplitudes for a particular $\lambda$ into a ratio of GPD to PDF:
\begin{align}
R (\lambda) &= \frac{\mbox{Im} {\cal T}_{{\small{\mbox{DVCS}}}}}{\mbox{Im} {\cal T}_{{{\mbox{DIS}}}}} = \frac{H^{S} (v=\xi\frac{1+\lambda}{1-\lambda},\xi)}{q^S(X)} \, , \nonumber\\
&=\frac{{\cal F}^{S}\left(X=\zeta/(1-\lambda),\zeta\right)}{(1-\zeta/2)q^{S}(X)} \, ,
\end{align}
\noindent i.e., 
\begin{equation}
{\cal F}^{S} \left(\zeta/(1-\lambda), \zeta\right) = (1-\zeta/2) ~R(\lambda) ~q^{S} (X) \, .
\label{ajmeq}
\end{equation}
There are several comments in order at this point: $\lambda$ is now bounded
from above through $\lambda \leq 1-\zeta$. This implies that the relationship
between the ratios in Eq.~(\ref{ajmeq}) is only strictly true for
$\lambda=0$ (i.e., for DVCS for which $Q^{'2} = 0$).  The case $\lambda \neq 0$
should be viewed as follows: for $Q^2\sim2~\mbox{GeV}^2$, there is still
the possibility of having more than one rung in the partonic ladder.
Probing the uppermost rung with a virtuality $Q^2$ reveals the
distribution in momentum fractions, in this case $X=\zeta$ i.e., $X-\zeta=0$
corresponding to $\lambda=0$.  The next rung and its distribution in
momentum fractions can be accessed by `emitting' a photon with
space-like virtuality (i.e., $Q^{'2} > 0$).  As $ Q^{'2}$ increases
and one goes further down the ladder to where $X \gg \zeta$, one approaches
the forward limit. If one keeps the interpretation of the $s$-channel
cut as being equal to the imaginary part of the `scattering' amplitude
for $Q^{'2} \neq 0$, which, in LO, is directly proportional to GPD/PDF at
$X \neq \zeta$ rather than at $X = \zeta$, then the ratio $R$ is a direct measure
of quark singlet GPD to the quark singlet PDF for $X \neq \zeta$.
However, this logic is valid only at LO. At NLO, the situation
radically changes since, first, the gluon directly enters into the
amplitude, secondly, the convolution of the coefficient function with
the GPD is no longer as trivial as in LO and thirdly,
the value of $\alpha_s$ at low $Q^2$ is quite different in LO and NLO.
Therefore such a simple relation as in Eq.~(\ref{ajmeq}) should and
can no longer be valid.
In order to keep the ansatz as simple as possible, we will only
require that the model GPDs, at least in LO, produce a ratio, $R$,
which is in broad agreement with the $R$ values obtained in the AJM
from Eq.~(\ref{ajmeq}). 

\begin{figure}  
\centering
\mbox{\epsfig{file=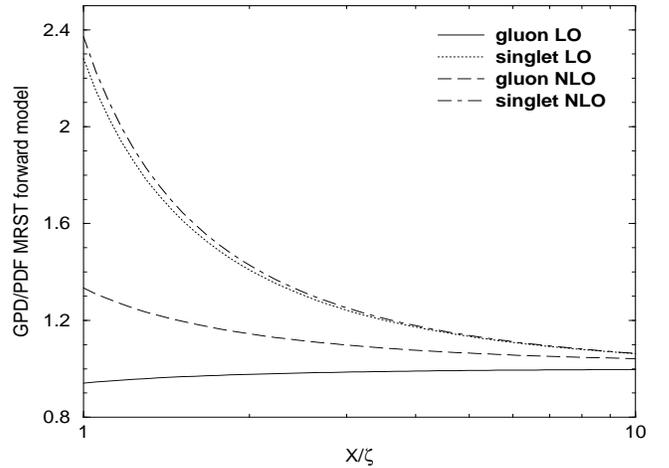,width=8.5cm,height=6.25cm}} 
\caption{The ratio GPD to PDF at $\zeta=0.0001$ for the quark singlet and gluon, using MRST01 distributions  in LO and NLO, at the input scale $Q_0 = 1$~GeV. This ratio is weakly dependent on $\zeta$, for small $\zeta$.}
\label{ratiof}
\end{figure}

If one chooses the forward model ansatz where
the GPD equals the PDF at $v$ in both LO and NLO, due to a lack of a
better Ansatz, (see e.g.  \cite{gb2}):
\begin{equation}
H^{S} (v,\xi) = q^S(v) \equiv q^S\left(\frac{X-\zeta/2}{1-\zeta/2}\right) \, ,
\label{fwd}
\end{equation}
\noindent which corresponds to the $b\to\infty$ limit of the DD model, 
one obtains a ratio of GPD to PDF at $X=\zeta$ of $\simeq 2.3$ for the quark
singlet, in agreement with the AJM prediction though for slightly
different values of $Q_0^2$ (directly compare the upper line in the
upper plot of Fig.~\ref{ajmratio} with the quark singlet in
Fig.~\ref{ratiof} keeping in mind that $\lambda=0 \Leftrightarrow X/\zeta=1$ and $\lambda=0.9 \Leftrightarrow
X/\zeta=10$). In consequence, our model ansatz corresponds to an AJM with
maximal skewedness.

For our forward model in the DGLAP region we, therefore, choose for
simplicity, the ansatz of Eq.~(\ref{fwd}) for the quark singlet, the
non-singlet (i.e., the valence) and the gluon.  The ratios of GPD to
PDF at the input scale for MRST01 quark singlet and gluon
distributions \cite{mrst01} at LO and NLO at $Q_0=1~\mbox{GeV}$ are
shown in Fig.~\ref{ratiof}.  For a function which falls as $x$
decreases, such as the valence quark at small $x$ or the MRST gluon at
LO, this ansatz leads to a suppression of the point $X = \zeta$ relative
to the forward case (see the lower, dotted line in Fig.~\ref{ratiof}).
Note that in NLO the MRST gluon actually goes negative at small $x$,
so a ratio of GPD/PDF $> 1 $ close to $X=\zeta$ in this case leads to a
suppression of the DVCS cross section from the gluon contribution,
relative to using PDFs. However, the DVCS cross section is rather
insensitive to the behavior of this ratio close to $X=\zeta$ for the
gluon, since it only enters in NLO, and is completely insensitive to
it for the non-singlet quark case since this distribution only enters
into the evolution.

The above reasoning indicates that the physics of the AJM model
provides a guide for modeling input GPDs in the DGLAP region.
If one compares the NLO imaginary part of the DVCS amplitude from the
above model to the NLO imaginary part of the DIS amplitude as
extracted from a recent H1 $F_2$ fit \cite{H1f2fit}, we find for
MRST2001, $R\simeq3.8$ and for CTEQ6M, $R\simeq2.7$ for $Q^2=3~\mbox{GeV}^2$ and
$x=0.0005$.  the enhancement effect of evolution, the AJM result
for $R$ is basically reproduced for CTEQ6M but not for MRST2001 using
eq.~(\ref{fwd}). Given the widely different parameterizations at NLO,
this seems acceptable to us, at present. As we will see below the
enhancement effect generated through the shift is too strong for both LO and
NLO at low values of $Q^2$ near the input scale.
  
The prescription in eq.~(\ref{fwd}) does not dictate what to do in the
ERBL region, which does not have a forward analog.  Naturally the GPDs
should be continuous through the point $X=\zeta$ and should have the
correct symmetries around the midpoint of the ERBL region. They are
also required to satisfy the requirements of polynomiality:
\begin{align}  
M_N & = \int^1_{-1} dv v^{N-1} [H^q (v,\xi) + H^g (v,\xi)] \nonumber \\ 
     &=  \sum^{N/2}_{k=0} \xi^{2k} C_{2k,N} \, . 
\label{polynomiality}  
\end{align}

At this point we choose to model the ERBL region with these natural
features in mind. We demand that the resultant GPDs reproduce the
first moment $M_1 \approx 3$ and the second moment $M_2 \approx 1$ \cite{footmom}
(with the D-term set to zero to remove the quadratic piece in
Eq.~(\ref{M2})) where
\begin{eqnarray}  
M_1 & = & \int^1_{0} dv H^{q,NS} (v,\xi)= 3 \, ,  \label{M1} \\
M_2 & = & \int^1_{0} dv v [H^{q,S} (v,\xi) + H^g (v,\xi)]\nonumber\\
& = & 1 + C~\xi^2 \label{M2}
\end{eqnarray}
and $C$ in Eq.~(\ref{M2}) was computed in the chiral-quark-soliton
model \cite{vanderhagen1}.  This reasoning suggests the following
simple analytical form for the ERBL region ($X < \zeta$):
\begin{align}
&{\cal F}^{g,NS}(X,\zeta) = {\cal F}^{g,NS}(\zeta) \left[ 1+A^{g,NS}(\zeta) C^{g,NS} (X,\zeta) \right] \, , \nonumber \\
&{\cal F}^{S}(X,\zeta) = {\cal F}^{S}(\zeta)\left(\frac{X - \zeta/2}{\zeta/2}\right) \left[1 + A^{S}(\zeta) C^{S} (X,\zeta) \right] \, ,
\label{ajmerbl}
\end{align}
\noindent where the functions  
\begin{align}
&C^{g,NS} (X,\zeta) = \frac{3}{2}\frac{2-\zeta}{\zeta}\left(1 - \left( \frac{X-\zeta/2}{\zeta/2} \right)^2 \right)  \, , \nonumber \\
&C^{S} (X,\zeta)  = \frac{15}{2}\left(\frac{2-\zeta}{\zeta}\right)^2 \left(1 - \left(\frac{X-\zeta/2}{\zeta/2} \right)^2 \right) \, ,
\end{align} 
\noindent vanish at $X=\zeta$ to guarantee continuity of the GPDs.
The $A^i(\zeta)$ are then calculated for each $\zeta$ by demanding that the
first two moments of the GPDs are explicitly satisfied
(remembering to include the D-term in the ERBL region which only
provides the quadratic term in $\xi$ in Eq.~(\ref{M1})). For the second
moment what we do in practice is to set the D-term to zero and demand
that for each flavor the whole integral over the GPD is equal to the
whole integral over the forward PDF for the input distribution
concerned (due to the inherent small errors on the PDFs, the sum of
such integrals will be close to, but not precisely equal to, unity).
Note that the modeling of the ERBL region is unimportant at small $\zeta$
since the unknown subtraction constant in the dispersion relation
between the real part of the amplitude which formally depends on both
the ERBL and the DGLAP region, and the imaginary part which formally
depends only on the DGLAP region, is proportional to $\zeta$ and therefore
inconsequential at small $\zeta$.

It would be straightforward to extend this algorithm to satisfy
polynomiality to arbitrary accuracy by writing the $A^i(\zeta)$ explicitly
as a polynomial in $\zeta$ where the first few coefficients are set by the
first two moments and the other coefficients are then randomly chosen
since nothing is known about them. The above algorithm is extremely
flexible both in its implementation and adaption to either other
forward PDFs or other functional forms in the ERBL region. Therefore
it can be easily incorporated into a fitting procedure.

In Fig.~\ref{fwdgpd} we show the shape of the resulting input GPDs for
two characteristic values of $\zeta = 0.001, 0.1$. The upper plot in this
figure explicitly shows the antisymmetry of the singlet GPD and the
symmetry of the gluon GPD about the point $X=\zeta/2$.
\begin{figure}  
\centering 
\mbox{\epsfig{file=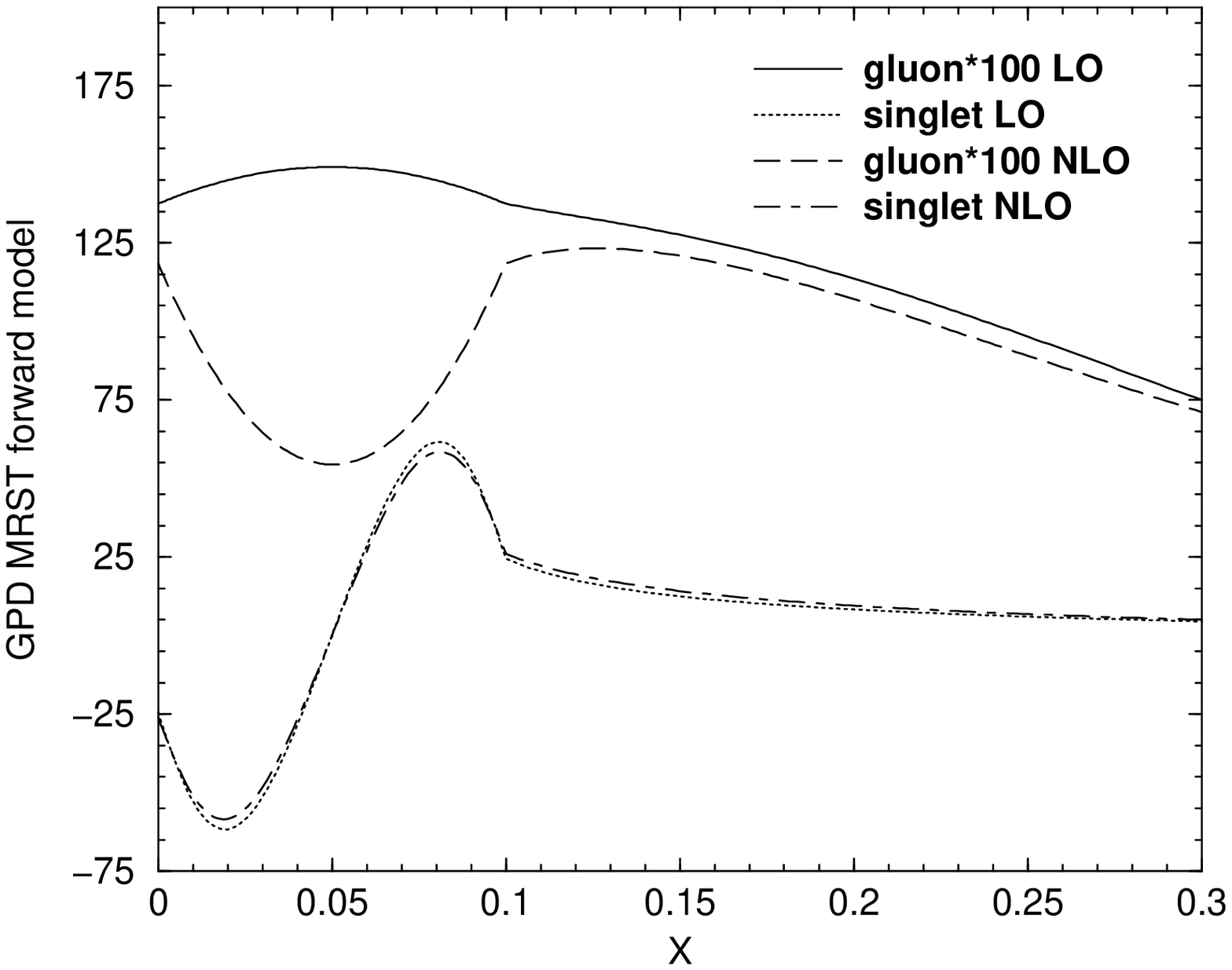,width=8.5cm,height=6.25cm}}
\end{figure}
\begin{figure}  
\centering
\mbox{\epsfig{file=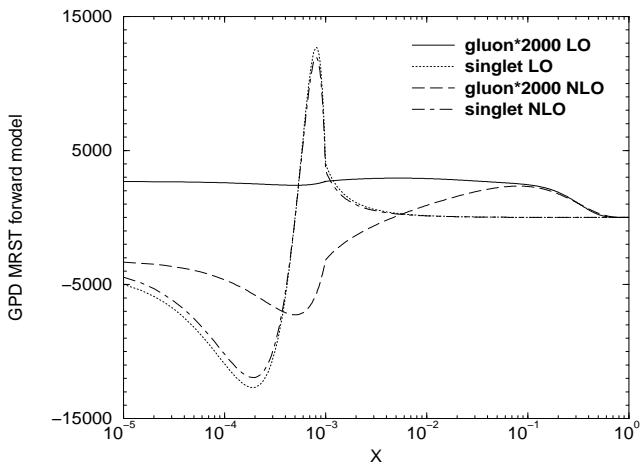,width=8.5cm,height=6.25cm}}
\caption{The quark singlet and gluon GPDs in LO and NLO, using MRST01 input PDFs, at 
the input scale $Q_0=1~\mbox{GeV}$ for $\zeta = 0.1$ (upper plot) and $\zeta = 0.001$ (lower plot), values 
typical of HERA and HERMES kinematics, respectively.}
\label{fwdgpd}
\end{figure}

\begin{figure}  
\centering 
\mbox{\epsfig{file=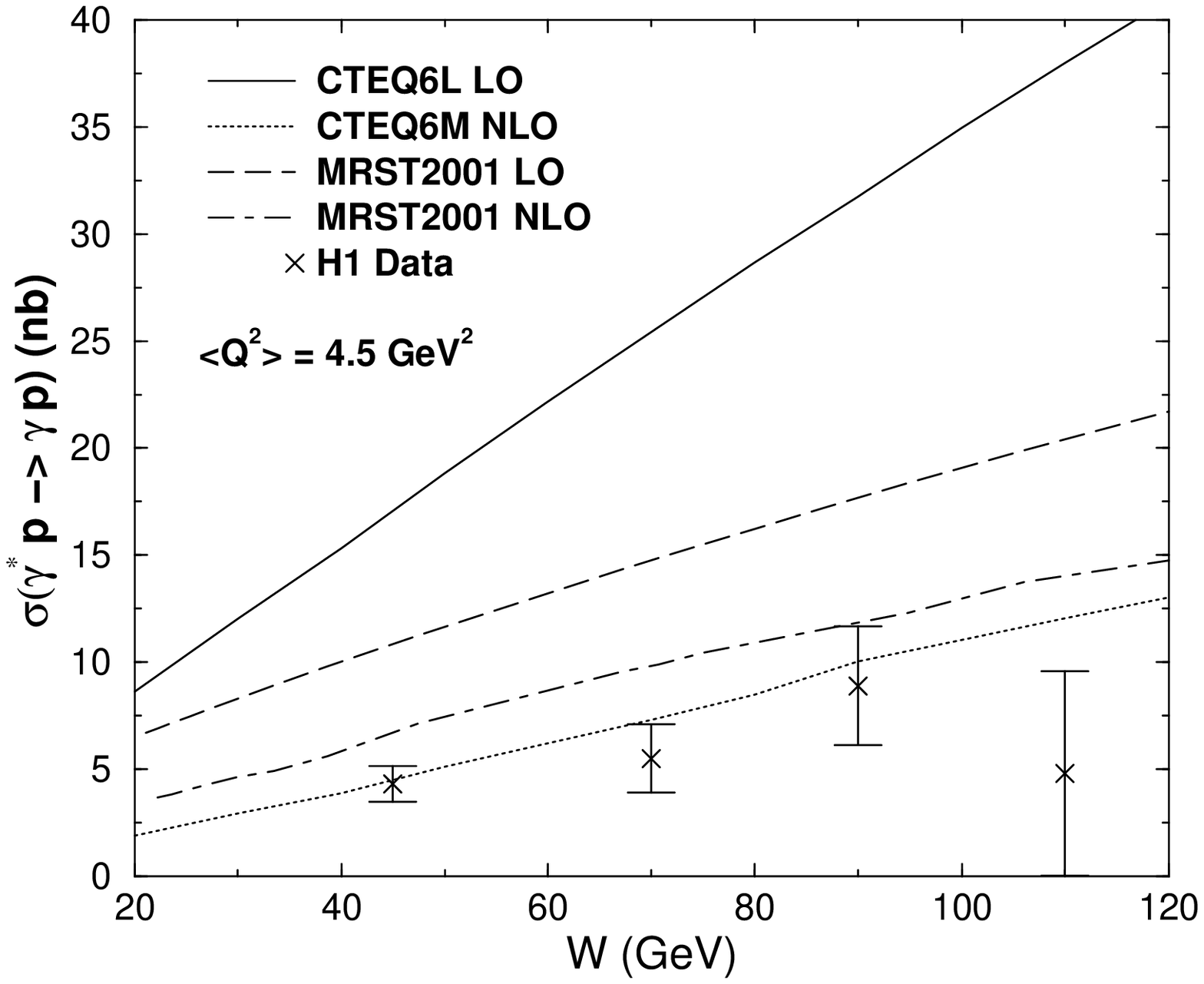,width=8.5cm,height=6.25cm}}
\mbox{\epsfig{file=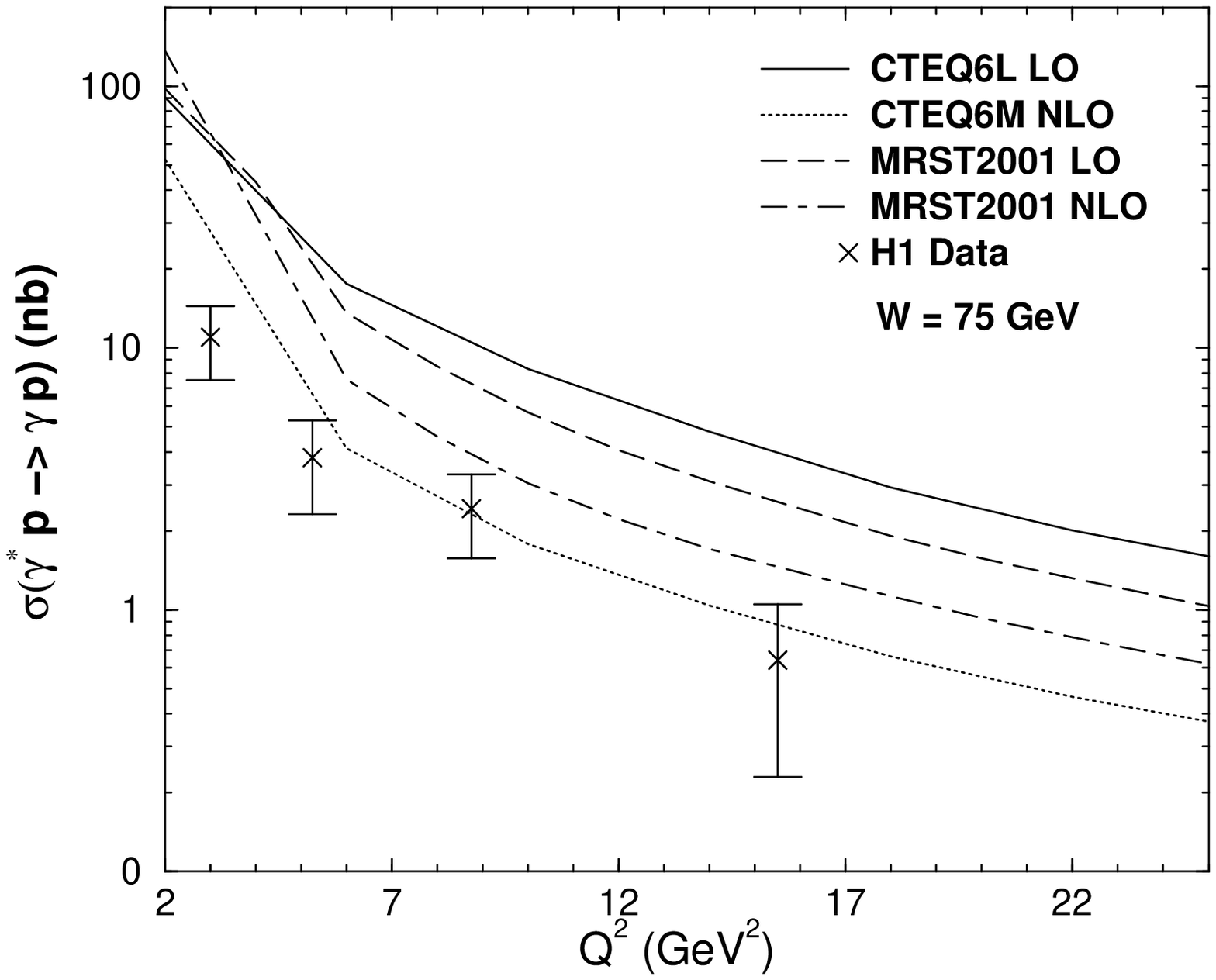,width=8.5cm,height=6.25cm}}
\caption{The photon level cross section, $\sigma (\gamma^* p \to \gamma p)$, calculated using 
the forward model ansatz for input GPDs, in the average kinematics of the H1 data: 
as a function of $W$ at fixed $Q^2 = 4.5$~GeV$^2$ (upper plot), and as a function of $Q^2$ 
at fixed $W=75~\mbox{GeV}$ (lower plot). A constant slope parameter of $B = 6.5$~GeV$^{-2}$ was used.}
\label{h1fig}
\end{figure}

\begin{figure}  
\centering 
\mbox{\epsfig{file=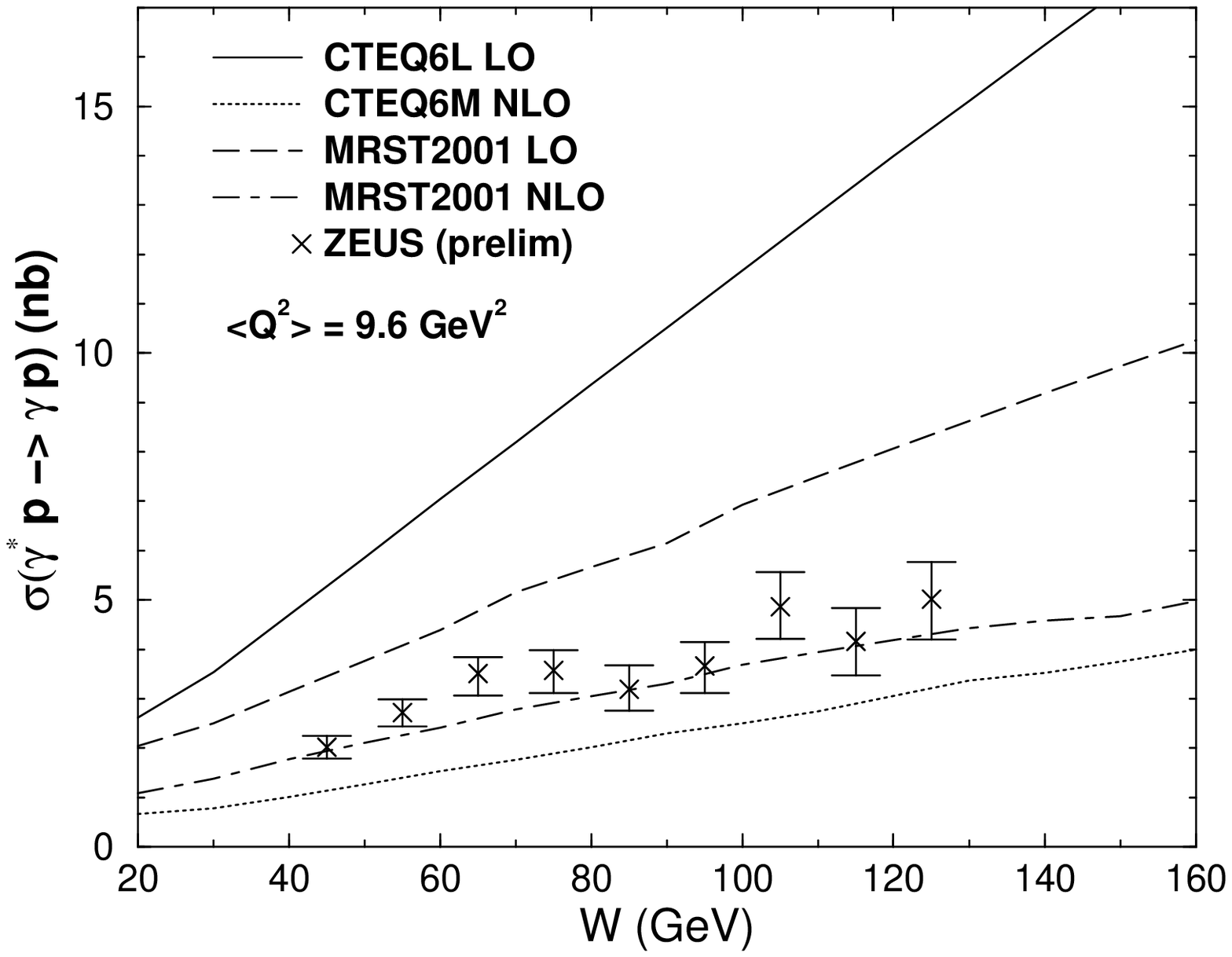,width=8.5cm,height=6.25cm}}
\mbox{\epsfig{file=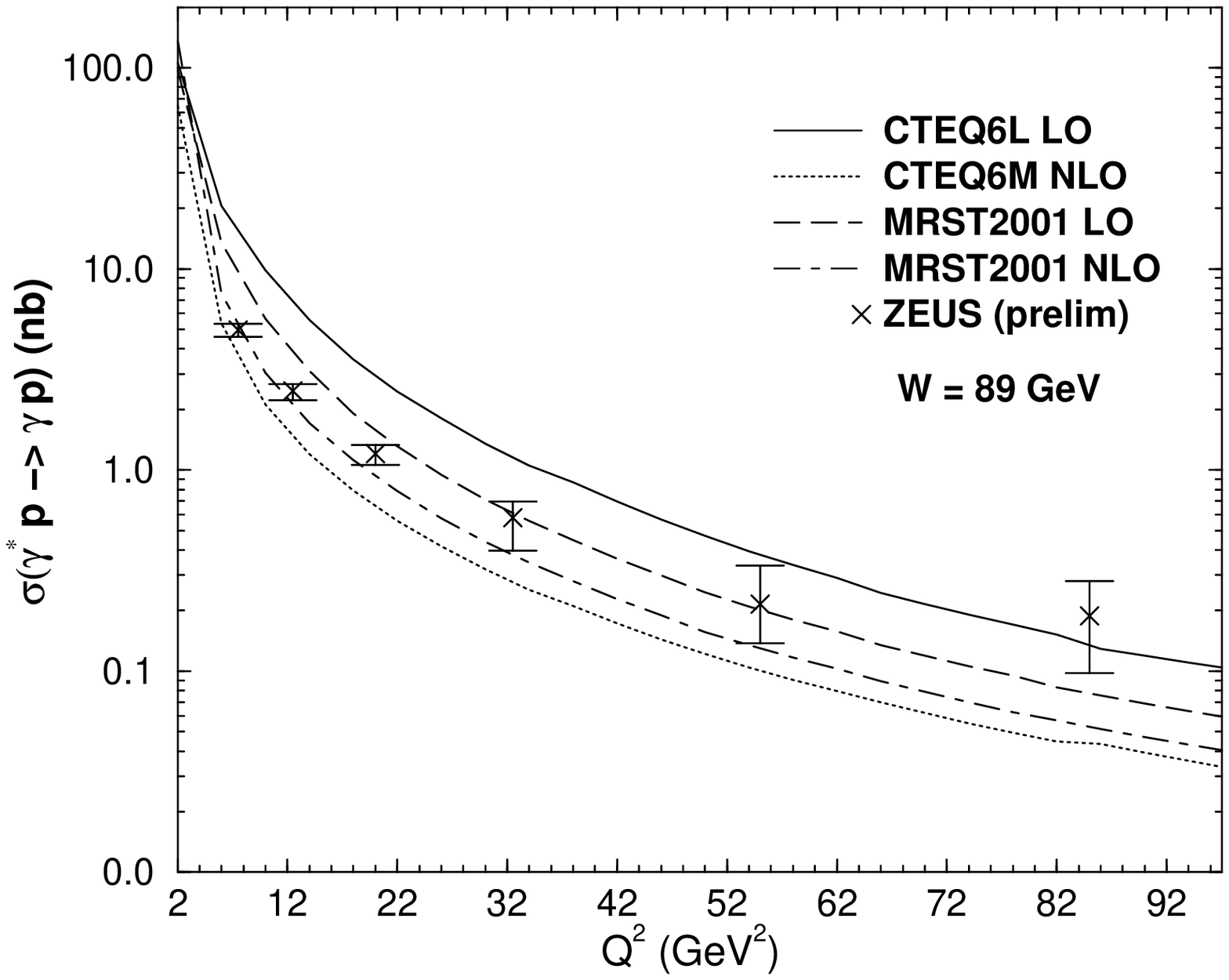,width=8.5cm,height=6.25cm}}
\caption{The photon level cross section, $\sigma (\gamma^* p \to \gamma p)$, calculated using 
the forward model ansatz for input GPDs, in the average kinematics of the preliminary ZEUS data: 
as a function of $W$ at fixed $Q^2 = 9.6$~GeV$^2$ (upper plot), and as a function of $Q^2$ 
at fixed $W=89~\mbox{GeV}$ (lower plot). A constant slope parameter of $B = 6.5$~GeV$^{-2}$ was used.}
\label{zeusfig}
\end{figure}

The photon level cross section results from this model, using MRST01
\cite{mrst01} and CTEQ6 \cite{cteq6} input distributions at LO and
NLO, are compared in Figs.~\ref{h1fig},\ref{zeusfig} to the H1
\cite{h1} and ZEUS \cite{zeus2} data at their average kinematic
points, respectively. In these curves we chose to use an $x$ and
$Q^2$-independent slope parameter of $B = 6.5$~GeV$^{-2}$, but
realistically there is a $30-40 \%$ uncertainty associated with the
value of this unknown parameter. The figures illustrate that within
the framework of the forward input model for GPDs the DVCS cross
section remains rather sensitive to the choice of input PDF and to the
accuracy with which the calculation is performed (i.e., LO or NLO).
It should be noted that the description at lowest H1 value of $Q^2$ is
bad. However, the enhancement we chose using eq.~(\ref{fwd}) at
$Q^2_0=1-1.69~\mbox{GeV}^2$ corresponds more to the AJM at
$Q^2=3~\mbox{GeV}^2$ (compare Figs.~\ref{ajmratio},\ref{ratiof}).
Hence it is not surprising that the description at low $Q^2$ is not
good suggesting that the shift in $X$ in eq.~(\ref{fwd}) should be
less at lower values of $Q_0$ . When increasing the input scale of
CTEQ6M, as done below, we find an appropriate reduction in the cross
section at low $Q^2$ much more in line with the low $Q^2$ data and the
AJM value.

It is important to note that the preliminary ZEUS data lies
systematically above the H1 data (see Fig.~11 of \cite{zeus2}).
Overall NLO seems to be doing better than LO, particularly on the
slope of the energy dependence.  It is fair to say that all of the
theory curves appear to have a $Q^2$-dependence that is too steep to
describe all of the data. We will return to this point in the next section.

The difference between the MRST and CTEQ curves at LO and NLO reflects
the relative size of the quark singlet and gluon distributions for
each set. It is possible that more precise data on DVCS may eventually
allow a discrimination between various input scenarios using NLO QCD.
For this to be realistic one would first need to pin down the
uncertainty associated with the slope by explicitly measuring the
$t$-dependence.

\begin{figure}  
\centering 
\mbox{\epsfig{file=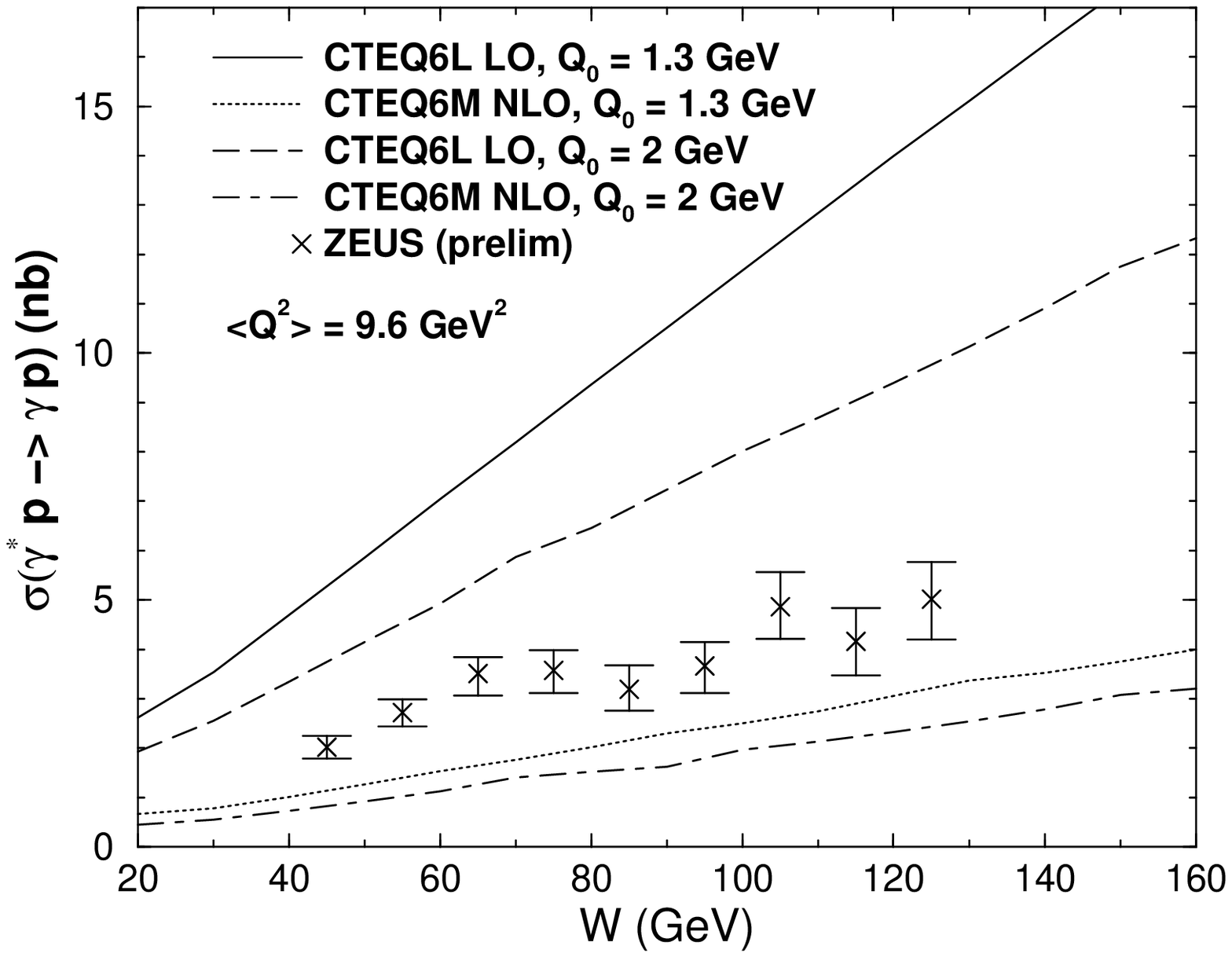,width=8.5cm,height=6.25cm}}
\mbox{\epsfig{file=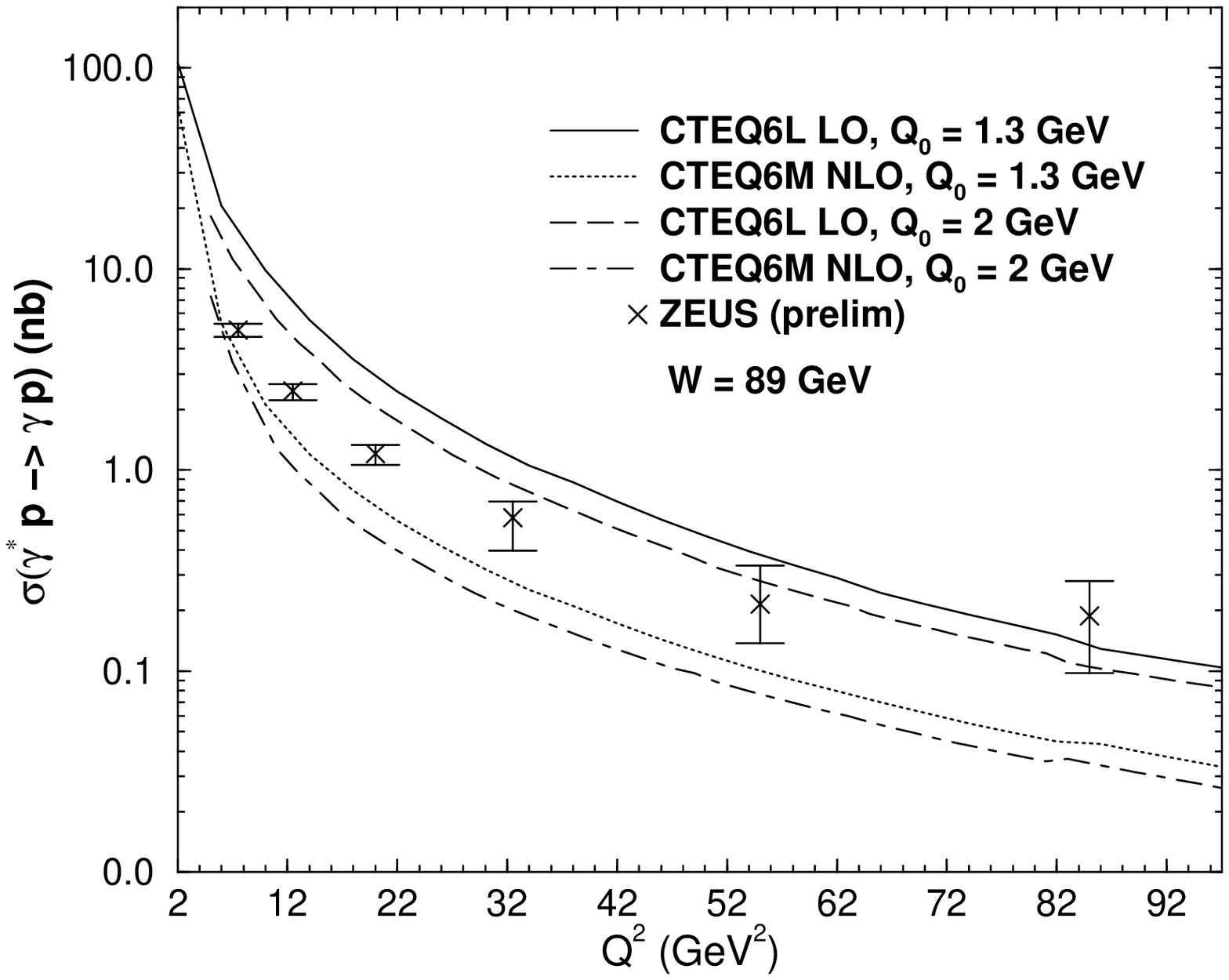,width=8.5cm,height=6.25cm}}
\caption{The effect of changing the starting scale, $Q_0$, on photon level 
  cross section $\sigma (\gamma^* p \to \gamma p)$, calculated using the forward model
  ansatz and CTEQ input PDFs, in the average kinematics of the
  preliminary ZEUS data: as a function of $W$ at fixed $Q^2 =
  9.6$~GeV$^2$ (upper plot), and as a function of $Q^2$ at fixed
  $W=89~\mbox{GeV}$ (lower plot). A constant slope parameter of $B =
  6.5$~GeV$^{-2}$ was used.}
\label{zeusfig2}
\end{figure}

We also investigated the effect on the cross section of increasing the
input scale for skewed evolution using CTEQ input distributions, from
the starting scale $Q_0 = 1.3$~GeV to $Q_0 = 2.0$~GeV. We then use the
forward PDFs at the new scale in our model for the GPDs.
Fig.~\ref{zeusfig2} shows that the reduced lever arm for skewed
evolution starting at the higher scale leads to a smaller cross
section at LO and NLO, as expected, and that, in LO at least, the
effect of this change is rather large.  In fact the CTEQ and MRST
collaborations only advocate the use of their forward PDFs above $Q^2
\approx 3-4$~GeV$^2$ (they start evolution at a lower scale $Q^2 \approx
1-2$~GeV$^2$ due to technicalities associated with a consistent
implementation of charm). Hence, it is not completely clear where one
should start skewed evolution, and this constitutes an additional
uncertainty in the theoretical predictions.

Having compared to small $x_{bj}$ data we will now test the AJM ansatz
for large $x_{bj}$ by comparing to data on the single-spin asymmetry,
SSA, (HERMES and CLAS \cite{hermclas}) and the charge asymmetry, CA,
(HERMES only), defined by
\begin{align}
&SSA = \frac{2 \int_0^{2\pi}d\phi ~\sin(\phi)(d\sigma^{\uparrow}-d\sigma^{\downarrow})}{\int_0^{2\pi}d\phi~(d\sigma^{\uparrow}+d\sigma^{\downarrow})} \, , \nonumber\\
&CA =  \frac{2\int_0^{2\pi}d\phi ~\cos(\phi)(d\sigma^{+}-d\sigma^{-})}{\int_0^{2\pi}d\phi~(d\sigma^{+}+d\sigma^{-})} \, . 
\label{defasym}
\end{align}
\noindent Here $d\sigma^{\uparrow}$, $d\sigma^{\downarrow}$ refer to the differential cross sections with the lepton 
polarized along or against its direction of motion, respectively;
$d\sigma^{+}, d\sigma^{-}$ are the unpolarized differential cross sections for
positrons and electrons, respectively.

Such a comparison of QCD models with the available high $x_{bj}$ data
may be viewed with some scepticism, especially in the case of CLAS
which has such a low $Q^2 \sim 1-2~\mbox{GeV}^2$ (HERMES is only slightly
better with a typical $Q^2$ of $\sim 2-4~\mbox{GeV}^2$).  Firstly, it is
{\it a priori} not clear that perturbation theory is applicable at
such low $Q^2$ values, (in particular, higher twist corrections may be
expected to become important in this region and our approximations
correspond to the DVCS cross section being divergent as $Q^2\to 0$.).
Secondly, the previously neglected GPDs $\tilde H,E$ and $\tilde E$
become increasingly important as $x$ increases \cite{afmmlong,bemu3}.
In the following we will include the dominant twist-3 contributions
\cite{bemu4}, which are entirely kinematic in origin, in our
calculation of the differential cross section, neglecting the
sub-dominant twist-3 effects.  We use the same input models for
$\tilde H,E$ and $\tilde E$ as well as $t$-dependence for the various
amplitudes i.e. various dipole form factors which are about
equivalent with an exponential at small $t$ with a slope $B$ between
$5$ and $8$, as in \cite{afmmlong}.  For HERMES we perform a full LO
and NLO QCD analysis, whereas for CLAS we are restricting ourselves to
LO, i.e., we are testing handbag dominance with no or little
evolution.  Furthermore, we shall restrict ourselves to MRST01 input
PDF for simplicity.

It transpires that the average kinematics of HERMES is such that $H$
is still the leading GPD and within our model assumptions $\tilde H,E$
and $\tilde E$ could be set to zero for those values, with negligible
difference to the final answer \cite{foot7b}.  Within the above
caveats, we find for average HERMES kinematics ($\langle x \rangle = 0.11, \langle Q^2 \rangle
= 2.56~\mbox{GeV}^2, \langle t \rangle = -0.265~\mbox{GeV}^2$)
\begin{align} 
\mbox{SSA} &= -0.28~\mbox{(LO)},~ ~-0.23~\mbox{(NLO)}\nonumber\\
\mbox{CA} &= 0.12~\mbox{(LO)},~ ~ 0.09~\mbox{(NLO)}
\end{align} 
compared to the quoted experimental results \cite{hermnew} 
\begin{align} 
\mbox{SSA} &= -0.21 \pm 0.08\nonumber\\
\mbox{CA} &= 0.11 \pm 0.07.
\end{align}
For the average CLAS
kinematics ($\langle x \rangle = 0.19, \langle Q^2 \rangle = 1.31~\mbox{GeV}^2, \langle t \rangle =
-0.19~\mbox{GeV}^2$) we find
\begin{equation} 
\mbox{SSA} = 0.2~\mbox{(LO)}
\end{equation}
compared to the experimental value (second reference of \cite{hermclas})
 \begin{equation} 
\mbox{SSA} = 0.202 \pm 0.041.
\end{equation}
This demonstrates that
the AJM ansatz works surprisingly well even at large
$x_{bj}$ giving us confidence in the AJM-based model and suggesting
that a fit to the available data should be possible without tuning too
many input parameters.

\section{A simple model for the slope parameter}
\label{slope}  

It was pointed out in \cite{ffs} that the $t$-slope of the DVCS cross
section at small $x$ should depend strongly on $Q^2$ in the
transitional region from $Q^2$ of a few GeV$^2$ to large $Q^2$. At
$Q^2 \sim 2$~GeV$^2$ it is natural to expect that the slope will be
pretty close to that for exclusive $\rho$-meson production: $B \sim
8$~GeV$^{-2}$ \cite{brho}. For large $Q^2$ the dominant contribution
is governed by evolution trajectories which, at the resolution $Q_0^2
\sim 2$~GeV$^2$, originate from the gluon field. Hence we expect that in
this case the slope will be given by the square of the two-gluon form
factor of the nucleon at $X, X-\zeta \gg x_{bj}$. Recently \cite{fs} it was
demonstrated that for $x_{bj} \geq 0.05$ this $t$-dependence can be
approximated in a wide range of $t$ as $ 1/(1-t/m^2_{2g})^4 $ with
$m^2_{2g} \sim 1.1$~GeV$^2$. This corresponds to a $t$-slope of $B \sim
3$~GeV$^{-2}$ for exponential fits \cite{foot8}.  At smaller $x$ an
increase of the slope is expected which could originate from several
effects, including Gribov diffusion. Hence for the highest $Q^2$ point
of ZEUS of about $90$~GeV$^2$ we expect $B = 3.5 \pm 0.5$~GeV$^{-2}$.
The recent H1 and ZEUS $\rho$-meson production data, for a $W$-range
similar to the DVCS experiments, indicates that the slope of
$\rho$-production both for $\sigma_L$ and $\sigma_T$ drops rather rapidly with
increasing $Q^2$ reaching $B \sim 5 $ at $Q^2 \approx 10$~GeV$^2$ \cite{brho2}.

\begin{figure}  
\centering 
\mbox{\epsfig{file=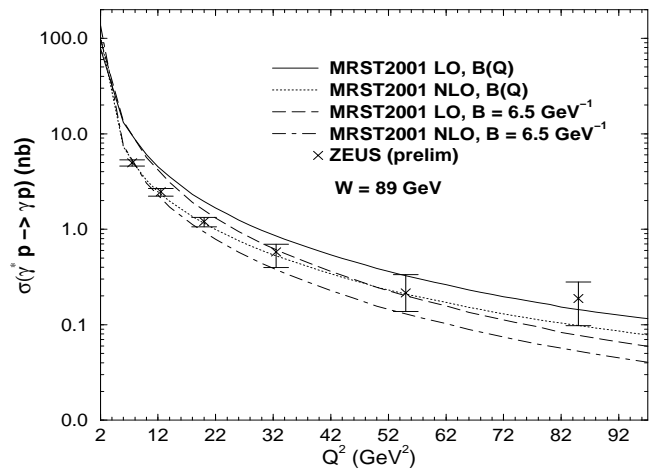,width=8.5cm,height=6.25cm}}
\caption{The effect on the DVCS cross section, in the average kinematics of the ZEUS data, 
of introducing our simple $Q^2$-dependent model of Eq.(\ref{bq}) for $B$, the slope of the $t$-dependence.}
\label{bfig}
\end{figure}

A simple parameterization which reflects the discussed constraints for
the range of $2\leq Q^2 \leq 100 $~GeV$^2$ is

\begin{equation}
B(Q^2) = B_0  \left(1 - C \ln \left(\frac{Q^2}{Q_0^2}\right)\right)  
\label{bq}
\end{equation}
\noindent with $B_0 = 8$~GeV$^{-2}$, $Q_0 = 2$~GeV$^2$, $C = 0.15$ being reasonable values for 
the various parameters. This gives $B (Q^2 = 9.6) = 6.1$~GeV$^{-2}$
and $B (Q^2 = 4.5) = 7.0$~GeV$^{-2}$ at the average $Q^2$ values of
the ZEUS and H1 data, respectively (in broad agreement with our chosen
constant value of $B = 6.5$~GeV$^{-2}$).  Fig. \ref{bfig} illustrates
the effect of introducing this simple model on the description of the
$Q^2$-dependence of the ZEUS data.

This modification of the $B$ slope gives a great improvement in
comparison with the data and shows how important an experimental
determination of the $B$ slope is, since it constitutes a large
theoretical uncertainty at this point.

\section{Conclusions}
\label{con}  

Using sound phenomenological guidelines such as the aligned jet model,
we have constructed a model for generalized parton distributions at
the input scale.  Within certain theoretical uncertainties (such as
the exact shape, the input scale and the functional form of $B$ in $W$
and $Q^2$) this model can be used in a NLO QCD analysis to describe
the recent DVCS data from H1, ZEUS, HERMES and CLAS within their
experimental errors.  In constructing this model we have given a
simple and flexible algorithm which can be easily incorporated into a
fitting procedure.

We have also demonstrated and explained the failure of the most widely
used model for generalized parton distributions, the factorized double
distribution based model, to describe the available DVCS data, when
rigorously applied in a LO or NLO QCD analysis in its current form.

The modeling of the input GPDs is now sufficiently advanced to
justify attempting to fit some of the input parameters directly to the
available data. A basic analysis of the data would seem to favor a
$t$-dependence with a slope parameter, $B$, that depends on $Q^2$.
Hence, an accurate measurement of this slope is of crucial importance
for further progress of the comparison of theory and experiment.

We would like to thank H.~Abramowicz, C.~Weiss, M.~Diehl,
L.~Frankfurt, K.~Golec-Biernat, D.~M{\"u}ller and A.~Sch{\"a}fer for helpful
discussions. A.F. was supported by the DFG Emmi-Noether Program, M.M.
by PPARC and M.S. by the DOE. This work is dedicated to Dina, Joanne
and Patricia.


\begin{thebibliography}{99} 
\bibitem{mrgdh} D.~M\"{u}ller {\it et al.}, Fortschr. Phys. {\bf 42}, 101 (1994).
\bibitem{ji} X.~Ji, Phys. Rev. D {\bf 55} 7114 (1997); J.\ Phys.\ G {\bf 24}, 1181 (1998). 
\bibitem{rad1} A.~V.~Radyushkin, Phys.\ Rev.\ D {\bf 56}, 5524 (1997). 
\bibitem{die} M.~Diehl {\it et al.}, Phys.\ Lett.\ B {\bf 411}, 193 (1997). 
\bibitem{mfort} D.~M{\"u}ller {\it et al.}, Fortsch.\ Phys.\ {\bf 42}, 101 (1994). 
\bibitem{vgg} M.~Vanderhaeghen, P.~A.~M.~Guichon and M.~Guidal, Phys.\ Rev.\ D {\bf 60}, 094017 (1999).
\bibitem{jcaf} J.~C.~Collins and A.~Freund, Phys.\ Rev.\ D {\bf 59}, 074009 (1999).
\bibitem{ffgs} L.~Frankfurt {\it et al.}, Phys.~Lett.~B {\bf 418}, 345 (1998), Erratum-ibid. {\bf 429}, 414 (1998); A.~Freund and V.~Guzey, Phys.~Lett.~B {\bf 462} 178 (1999).  
\bibitem{ffs} L.~Frankfurt, A.~Freund and M.~Strikman, Phys.\ Rev.\ D {\bf 58}, 114001 (1998), Erratum-ibid. D {\bf 59}, 119901 (1999); Phys.\ Lett.\ B {\bf 460}, 417 (1999);  A.~Freund and M.~Strikman, Phys.\ Rev.\ D {\bf 60}, 071501 (1999).
\bibitem{gb1} K.~J.~Golec-Biernat and A.~D.~Martin, Phys.\ Rev.\ D {\bf 59}, 014029 (1999). 
\bibitem{gb2} A.~G.~Shuvaev {\it et al.}, Phys. Rev. D {\bf 60}, 014015 (1999);  K.~J.~Golec-Biernat, A.~D.~Martin and M.~G.~Ryskin, Phys.~Lett.~B {\bf 456}, 232 (1999).
\bibitem{bemuothers}  A.~V.~Belitsky, D.~M\"{u}ller, Phys.~Lett.~B{\bf 417}, 129 (1998), Nucl.~Phys.~B{\bf 537}, 397 (1999),  Phys.~Lett.~B{\bf 464}, 249 (1999), Nucl.~Phys.~B{\bf 589}, 611 (2000), Phys.~ Lett.~B{\bf 486}, 369 (2000), A.~V.~Belitsky, A.~ Freund, D.~M\"{u}ller, Phys.~Lett.~B{\bf 461}, 270 (1999), Nucl.~Phys.~B{\bf 574}, 347 (2000), Phys.~Lett.~B{\bf 493}, 341 (2000).     
\bibitem{foot} Regular parton distributions are merely particle distributions not particle correlation functions. 
\bibitem{poly} M.~V.~Polyakov and C.~Weiss, Phys.\ Rev.\ D {\bf 60}, 114017 (1999).
\bibitem{factor} Factorized refers here to writing the DD, a three variable function, in terms of the product of a two variable with two one variable functions as explicitly shown in Sec.~\ref{ddfailure}.
\bibitem{rad2} A.~V.~Radyushkin, Phys.~Rev.~D {\bf 59}, 014030 (1999). 
\bibitem{rad3} A.~V.~Radyushkin, Phys.~Lett.~B {\bf 449}, 81 (1999).
\bibitem{h1} H1 Collaboration, C. Adloff {\it et al.}, Phys.\ Lett.\ B {\bf 517}, 47 (2001).
\bibitem{hermclas} HERMES Collaboration, A.~Airapetian {\it et al.}, Phys.~Rev.~Lett.\ {\bf 87}, 182001 (2001);
CLAS Collaboration, S.~Stepanyan {\it et al.}, Phys.~Rev.~Lett. {\bf 87}, 182002 (2001);
\bibitem{afmmshort} A.~Freund and M.~McDermott, Phys.~Rev.~D {\bf 65}, 091901 (2002).
\bibitem{afmmlong} A.~Freund and M.~McDermott, Eur.~ Phys.~ J.~C {\bf 23}, 651 (2002). 
\bibitem{bemuk1} A.~V.~Belitsky {\it et al.}, Nucl.~ Phys.~B {\bf 629}, 323 (2002).
\bibitem{ajm}  J.\ D.\ Bjorken {\it et al.} Phys.~ Rev.~D {\bf 8} 1341 (1973).
\bibitem{fs88} L.~L.~Frankfurt and M.~I.~Strikman,
Phys.\ Rept.\  {\bf 160}, 235 (1988); Nucl.\ Phys.\ B {\bf 316}, 340 (1989).
\bibitem{afmmgpd} A.~Freund and M.~McDermott, Phys.~Rev.~D {\bf 65}, 074008 (2002).
\bibitem{bemuevolve} A.~V.~ Belitsky, D.~ M{\"u}ller, L.~ Niedermeier, A.~ Sch{\"a}fer, Nucl.~Phys.~B{\bf 546},279 (1999),  Phys.~Lett.~B{\bf 437},160 (1998), A.~V.~ Belitsky, B.~Geyer, D.~ M{\"u}ller, A.~ Sch{\"a}fer, Phys.~Lett.~B{\bf 421}, 312 (1998).
\bibitem{radmus} I.~V.~Musatov and A.~V.~Radyushkin, 
Phys.~Rev.~D {\bf 61}, 074027 (2000).   
\bibitem{footpi} The fact that $\pi(x,y) = \pi(x,-y)$ guarantees that odd powers 
of $\xi$ are missing from the moments of $H(v,\xi)$. This is required by 
hermiticity and time reversal of the corresponding hadronic matrix elements.
\bibitem{foot3} Comparison of any of the above models with data from HERMES 
and CLAS shows that the same problem persists also for large $x_{bj}$, as 
already pointed out in \cite{afmmshort}.
\bibitem{bemu2} A.~V.~Belitsky {\it et al.}, Phys.~Lett.~B {\bf 510}, 117 (2001). 
\bibitem{afmmamp} A.~Freund and M.~McDermott, Phys.~Rev.~D {\bf 65}, 056012 (2002).
\bibitem{bemus} A.~V.~ Belitsky, D.~ M{\"u}ller, L.~ Niedermeier, A.~ Sch{\"a}fer, Phys.~ Lett.~ B{\bf 474}, 163 (2000).  
\bibitem{grv} M.~Gl{\"u}ck, E.~Reya and A.~Vogt, Eur.~Phys.~J.~C {\bf 5}, 461 (1998). 
\bibitem{zeus} ZEUS Collaboration, P.~R.~Saull, `Prompt photon production and observation of deeply virtual Compton scattering', Proc. ICHEP 1999, (Tampere, Finland, July 1999), hep-ex/0003030.  
\bibitem{zeus2} ZEUS Collaboration, `Measurement of the deeply virtual Compton scattering cross section at HERA', contributed paper (abstract: 825) to ICHEP2002 (Amsterdam, July 2002).
\bibitem{mrst01}  A.~D.~Martin {\it et al.}, Eur.~Phys.~J.~C {\bf 23}, 73 (2002).
\bibitem{footmom} The first moment counts the number of quarks in the proton 
and the second moment is a generalization of the momentum sum rule where the D-term generates all the deviation from unity as $\xi$ varies.
\bibitem{vanderhagen1} N.~Kivel, M.~V.~Polyakov and M.~Vanderhaeghen, Phys.~Rev.~D {\bf 63}, 114014 (2001).
\bibitem{cteq6} J. Pumplin {\it et al.}, CTEQ Collaboration, J. High Energ. Phys. {\bf 0207}, 012 (2002).
\bibitem{bemu3} A.~V.~Belitsky {\it et al.}, Nucl.~Phys.~B {\bf 593}, 289 (2001).  
\bibitem{bemu4} A.~V.~Belitsky, D.~M\"{u}ller and A.~Kirchner, Nucl.~Phys.~B {\bf 629}, 323 (2002).
\bibitem{H1f2fit} H1 Collaboration, Eur.~Phys.~J.~C {\bf 21}, 33 (2001).
\bibitem{foot7b} At very small $t$, $\tilde E$ becomes the large due to the pion pole it contains. However, due to other kinematical factors in front of it, the contribution from the term involving $\tilde E$ only becomes important close to $t_{min}$, which is never 
reached at either HERMES or CLAS.
\bibitem{hermnew} HERMES Collaboration, Nucl.~Phys.~ A {\bf 711}, 171 (2002).
\bibitem{brho} ZEUS Collaboration, `Exclusive and proton-dissociative
  electroproduction of $\rho^{0}$ mesons at HERA', contributed paper
  (abstract: 818) to ICHEP2002 (Amsterdam, July 2002).  
\bibitem{fs} L.~Frankfurt and M.~Strikman, Phys.~Rev.~D {\bf 66}, 031502 (2002).  
\bibitem{foot8} One needs to investigate how a change of the $t$-slope parameterization in the Monte Carlo used to analyse the ZEUS and H1 data would modify the extracted cross section.  
\bibitem{brho2} H1 Collaboration, `Elastic electroproduction of $\rho$ mesons at high $Q^2$ at HERA', contributed paper (abstract: 989) to ICHEP2002 (Amsterdam, July
  2002).

\end{thebibliography}
\end{document}